\documentclass[twocolumn,pra,superscriptaddress]{revtex4}
\usepackage{amsmath}
\usepackage{graphicx}
\usepackage{amssymb}
\usepackage{dcolumn}
\usepackage{mathrsfs}
\usepackage{bm}
\usepackage{color}
\makeatletter

\newcommand{\Rmnum}[1]{\expandafter\@slowromancap\romannumeral #1@}
\makeatother

\begin{document}

\title{Emergent pseudospin-1 Maxwell fermions with threefold degeneracy in optical lattices}

\author{Yan-Qing Zhu}
\affiliation{National Laboratory of Solid State Microstructures
and School of Physics, Nanjing University, Nanjing 210093, China}

\author{Dan-Wei Zhang}
\email{zdanwei@126.com}\affiliation{Guangdong Provincial Key Laboratory of Quantum Engineering and Quantum Materials,
SPTE, South China Normal University, Guangzhou 510006, China}

\author{Hui Yan}
\affiliation{Guangdong Provincial Key Laboratory of Quantum
Engineering and Quantum Materials, SPTE, South China Normal
University, Guangzhou 510006, China}

\author{Ding-Yu Xing}
\affiliation{National Laboratory of Solid State Microstructures
and School of Physics, Nanjing University, Nanjing 210093, China}
\affiliation{ Collaborative Innovation Center of Advanced
Microstructures, Nanjing 210093, China}

\author{Shi-Liang Zhu}
\email{slzhu@nju.edu.cn}
\affiliation{National Laboratory of Solid
State Microstructures and School of Physics, Nanjing University,
Nanjing 210093, China}

\affiliation{Guangdong Provincial Key Laboratory of Quantum
Engineering and Quantum Materials, SPTE, South China Normal
University, Guangzhou 510006, China}

\affiliation{Synergetic Innovation Center
of Quantum Information and Quantum Physics, University of Science
and Technology of China, Hefei, Anhui 230026, China}

\date{\today}

\begin{abstract}
The discovery of relativistic spin-1/2 fermions such as Dirac and
Weyl fermions in condensed matter or artificial systems opens a
new era in modern physics. An interesting but rarely explored
question is whether other relativistic spinal excitations could be
realized with artificial systems. Here, we construct two- and three-dimensional
tight-binding models realizable with cold fermionic atoms in optical lattices,
where the low energy excitations are effectively described by the spin-1 Maxwell equations in the Hamiltonian form.
These relativistic (linear dispersion) excitations with unconventional integer pesudospin, beyond the Dirac-Weyl-Majorana
fermions, are a new kind of fermions named as Maxwell fermions. We demonstrate that the systems have rich topological features. For instance,
the threefold degenerate points called Maxwell points may have quantized Berry phases and anomalous quantum Hall effects with spin-momentum locking may appear in topological Maxwell insulators in the two-dimensional lattices. In three dimensions, Maxwell points may have nontrivial monopole charges of $\pm 2$
with two Fermi arcs connecting them, and the merging of the Maxwell points leads to topological phase transitions. Finally, we propose realistic schemes for
realizing the model Hamiltonians and detecting the topological properties of the emergent Maxwell quasiparticles in optical lattices.
\end{abstract}

\maketitle

\section{Introduction}

Discovery of new particles in nature or new
quasiparticles in condensed matter systems is at the heart of
modern physics \cite{Wilczek}. One of the remarkable examples is
the discovery of relativistic Dirac fermions emerged from
graphene, which has attracted great interest in physics
\cite{Castro}. Furthermore, it was demonstrated that Weyl
fermions, which are massless spin-1/2 particles according to
quantum field theory and never before observed as fundamental
particles in nature, can emerge as quasiparticles in condensed
matter \cite{Wan,Huang,Weng,Xu1,Lv1,Xu2,Lv2} or photonic crystals
\cite{Lu,Chan}. Most interestingly, Dirac and Weyl fermions have
rich topological features
\cite{Castro,Wan,Huang,Weng,Xu1,Lv1,Xu2,Lv2,Lu,Chan,Hasan,Qi}. However,
quasiparticles with higher spin numbers are also fundamentally
important but rarely studied in condensed matter physics or
artificial systems \cite{Bradlyn,Lan,Liang}. Recently, a pioneer
work in this direction theoretically predicted that "new fermions" (fermionic quasiparticle excitations)
beyond the Dirac-Weyl-Majorana classification can emerge in some
band structures with three- or more-fold degenerate points in the
presence of time-reversal symmetry \cite{Bradlyn}. Very recently, the observation of three-component new fermions in the
topological semimetal molybdenum phosphide has been reported \cite{Lv2017}.
Although it may be difficult to find these (and other) new fermions in
condensed matter systems, they may also emerge from well-designed and
tunable ultracold atomic systems, especially since Dirac and Weyl
fermions have already been well studied in the field of cold atoms
\cite{Zhu,Zhang2012,Tarruell,Duca,Bermudez,Dubcek,Zhang2015}.

In this paper, we propose and analyze an exotic kind of
pseoduspin-1 fermions in two-dimensional (2D) square and three-dimensional (3D) cubic optical lattices, dubbed ``Maxwell fermions" as they are analogous to massless spin-1 photons described by the Maxwell equations. We first rewrite the Maxwell equations in an
anisotropic medium in the form of the Schr\"{o}dinger equation and
then construct 2D and 3D tight-binding models, where the low-energy excitations are
described by the Maxwell Hamiltonian. With a tunable parameter,
the systems can be varied in different quantum phases: topological or
normal Maxwell insulator, and topological or normal Maxwell
metal. The topological Maxwell metal is characterized by threefold
degenerate points, known as Maxwell points, which have nontrivial
monopole charges or quantized Berry phases and the low-energy
excitation near the Maxwell point behaved like a photon described
by the Maxwell equations. In the 2D system, we find
interesting spin-momentum locking edge states in the Maxwell
insulating phase, which is in analogy with the circularly-polarized polarization of the photons. In the 3D
system, the topological properties of Maxwell fermions are similar to those of the Weyl fermions in Weyl
semimetals, and the Maxwell points have monopole charges of $\pm 2$ with two Fermi arcs connecting them. The experimental scheme for implementation (and detection) of our models using three-component ultracold atoms in optical lattices is presented. Although some threefold band degeneracies
were revealed in solid-state systems with the body-centered
lattice structure and time-reversal symmetry in Ref.
\cite{Bradlyn}, our proposal is essentially different. First, our
threefold Maxwell points in 3D exist when the time-reversal symmetry of the system
is broken, in which case the threefold band degeneracies in Ref.
\cite{Bradlyn} will split into a number of Weyl points. Thanks to
the broken time-reversal symmetry, the minimal number of threefold
degeneracies in our 3D system can be two (which is thus a minimal
model), instead of four in Ref. \cite{Bradlyn}. In addition, we use
three atomic internal states to form the pseudospin-1 basis and
thus only the simple cubic lattice is used in our proposal, in
contrast to the required body-centered cubic lattices for spin-1/2
electrons in real materials. This enables us to realize exotic
threefold fermions in a lattice of simplest geometry.

The paper is organized as follows. In Sec. II, we rewrite the Maxwell equations in the Schr\"{o}dinger's form and then present the general idea of realizing Maxwell fermionic quasiparticles in lattice systems. Section III introduces the 2D square-lattice model for realizing the Maxwell metals and insulators, and explores the topological properties of the emergent Maxwell fermions. In Sec. IV, we proceed to study the properties of 3D Maxwell fermions in the cubic-lattice model. In Sec. V, we propose schemes for realization of the model Hamiltonian and detection of the topological Chern numbers in the optical lattices. Finally, a brief discussion and a short conclusion are given in Sec. VI.

\section{Maxwell Hamiltonian and Maxwell fermions in lattice systems}

In this section, we first rewrite the Maxwell equations in an
anisotropic medium in the form of the Schr\"{o}dinger equation, and
then describe the general scheme for realizing the Maxwell fermions in artificial lattice systems.

\subsection{Maxwell equations in the Schr\"{o}dinger's form}

In a region absent of charges and currents, the
well-known Maxwell equations in matter are given by
\begin{equation}\label{Maxwell}
\begin{split}
\nabla\times\mathbf{E}&=-\frac{\partial{\mathbf{B}}}{\partial{t}},\quad
\nabla\cdot\mathbf{{E}}=0,\\
\nabla\times\mathbf{H}&=\ \
\frac{\partial{\mathbf{D}}}{\partial{t}},\quad\
\nabla\cdot\mathbf{B}=0,
\end{split}
\end{equation}
where the displacement field
$\mathbf{D}=\varepsilon_0\varepsilon_r\mathbf{E}$ with
$\mathbf{E}$ being the electric field, and the magnetic field
$\mathbf{B}=\mu_0\mu_r\mathbf{H}$ with $\mathbf{H}$ being the
magnetizing field. Here $\varepsilon_0$ ($\mu_r$) is the
permittivity (permeability) of free space, and $\varepsilon_r$ and
$\mu_r$ are the relative permittivity and permeability,
respectively. In an anisotropic medium, $\varepsilon_r$ and
$\mu_r$ are tensors rather than numbers. To simplify the
proceeding analysis, we assume that the tensors $\varepsilon_r$
and $\mu_r$  are simultaneously diagonalized, then the
relationships between $\mathbf{D}$ and $\mathbf{E}$, $\mathbf{B}$
and $\mathbf{H}$ now become
\begin{equation}
\begin{split}
\left(\begin{matrix}D_x\\D_y\\D_z\end{matrix}\right)&=\varepsilon_0\left(\begin{matrix}\varepsilon_x&0&0\\0&\varepsilon_y&0\\0&0&\varepsilon_z\end{matrix}\right)\left(\begin{matrix}E_x\\E_y\\E_z\end{matrix}\right),\quad \\
\begin{pmatrix}B_x\\B_y\\B_z\end{pmatrix}&=\mu_0\begin{pmatrix}\mu_x&0&0\\0&\mu_y&0\\0&0&\mu_z\end{pmatrix}\left(\begin{matrix}H_x\\H_y\\H_z\end{matrix}\right).
\end{split}
\end{equation}
Thus Eq. (\ref{Maxwell}) can be rewritten as
\begin{equation}
\begin{aligned}
\epsilon_{\alpha\beta\gamma}\frac{\partial{E_\gamma}}{\partial\beta}
=-\frac{\partial{B_\alpha}}{\partial{t}}
&\Rightarrow\frac{c}{\sqrt{\varepsilon_\gamma\mu_\alpha}}\epsilon_{\alpha\beta\gamma}\frac{\partial{\tilde{E}_\gamma}}{\partial\beta}
=-\frac{\partial{\tilde{H}_\alpha}}{\partial{t}},\\
\epsilon_{\alpha\beta\gamma}\frac{\partial{H_\gamma}}{\partial\beta}=\
\frac{\partial{D_\alpha}}{\partial{t}}
&\Rightarrow\frac{c}{\sqrt{\varepsilon_\alpha\mu_\gamma}}\epsilon_{\alpha\beta\gamma}\frac{\partial{\tilde{H}_\gamma}}{\partial\beta}=\
\frac{\partial{\tilde{E}_\alpha}}{\partial{t}},
\end{aligned}
\end{equation}
where
$\tilde{E}_\alpha=\sqrt{\varepsilon_0\varepsilon_\alpha}E_\alpha$,
$\tilde{H}_\alpha=\sqrt{\mu_0\mu_\alpha}H_\alpha$, and
$c=1/\sqrt{\varepsilon_0\mu_0}$. Then we define the  photon wave
function as \cite{Oppenheimer,Good}
$$\mathbf{\Phi}(\mathbf{r},t)=\mathbf{\tilde{E}}(\mathbf{r},t)+i\mathbf{\tilde{H}}(\mathbf{r},t),$$
we have $\nabla\cdot\mathbf{\Phi}=0$, and
\begin{equation}
\begin{split}
i\hbar\frac{\partial{\Phi_m^\alpha}}{\partial{t}}=\nu_{\alpha\gamma}(i\epsilon_{\alpha\beta\gamma})\frac{\hbar}{i}\frac{\partial{\tilde{E}_\gamma}}{\partial\beta}
+i\nu_{\gamma\alpha}(i\epsilon_{\alpha\beta\gamma})\frac{\hbar}{i}\frac{\partial{\tilde{H}_\gamma}}{\partial\beta}.
\end{split}
\end{equation}
where $\nu_{\alpha\gamma}=c/\sqrt{\varepsilon_\alpha\mu_\gamma}$,
$\nu_{\gamma\alpha}=c/\sqrt{\varepsilon_\gamma\mu_\alpha}$,
$\hat{P}_\beta=-i\hbar\partial_{\beta}$.  When
$\varepsilon_\alpha\mu_\gamma=\varepsilon_\gamma\mu_\alpha$, that
is, $\nu_{\alpha\gamma}=\nu_{\gamma\alpha}$ [the condition for
obtaining a hermitian Hamiltonian, see Eq. (6)], then we can
further rewrite Eq. (4) as
\begin{equation}
\begin{split}
i\hbar\frac{\partial{\Phi_m^\alpha}}{\partial{t}}=\nu_{\alpha\gamma}(i\epsilon_{\alpha\beta\gamma})\hat{P}_\beta\Phi_m^\gamma.
\end{split}
\end{equation}
We hence obtain the following Schr\"{o}dinger's equation
\begin{equation}
\begin{split}
i\hbar\frac{\partial}{\partial{t}}\begin{pmatrix}\Phi_m^x\\
\Phi_m^y\\ \Phi_m^z\end{pmatrix}=
\begin{pmatrix}0&-i\nu_{xy}\hat{P}_z&i\nu_{xz}\hat{P}_y\\i\nu_{yx}\hat{P}_z&0&-i\nu_{yz}\hat{P}_x\\ -i\nu_{zx}\hat{P}_y&i\nu_{zy}\hat{P}_x&0\end{pmatrix}
\begin{pmatrix}\Phi_m^x\\ \Phi_m^y\\ \Phi_m^z\end{pmatrix}.
\end{split}
\end{equation}
This corresponds to the Maxwell equations in the anisotropic medium in the Schr\"{o}dinger's form
\begin{equation}
i\hbar\frac{\partial}{\partial{t}}\mathbf{\Phi}=\hat{H}_{M}\mathbf{\Phi},
\end{equation}
where the Hamiltonian is given by
\begin{equation}\label{MaxwellEq}
\hat{H}_{M}=v_x\hat{S}_x\hat{P}_x+v_y\hat{S}_y\hat{P}_y+v_z\hat{S}_z\hat{P}_z.
\end{equation}
Here
$\hat{S}_\beta=(\hat{S}_{\alpha\gamma})^\beta=i\epsilon_{\alpha\beta\gamma}$,
and $\epsilon_{\alpha\beta\gamma}$ $(\alpha,\beta,\gamma=x,y,z)$
is the Levi-Civita symbol. Noted that, $v_x=\nu_{yz}=\nu_{zy}$,
$v_y=\nu_{zx}=\nu_{xz}$, and $v_z=\nu_{xy}=\nu_{yx}$ are the
necessary and sufficient condition to obtain a hermitian
Hamiltonian in Eq. (6). Typical case when $\varepsilon_r=\mu_r=1$,
it returns to the free space situation and we obtain the related
Hamiltonian of single photon in vacuum as
$\hat{H}_{M}=c\hat{\mathbf{S}}\cdot\hat{\mathbf{P}}$.
 Here $\hat{\mathbf{S}}=(\hat{S}_x,\hat{S}_y,\hat{S}_z)$ are the
spin matrices for a particle of spin-1, which are defined as
\begin{equation}
\begin{split}
\hat{S}_x=\hat{S}^{1}= i\left(\begin{matrix}
    \varepsilon_{111}&\varepsilon_{112}&\varepsilon_{113}\\
    \varepsilon_{211}&\varepsilon_{212}&\varepsilon_{213}\\
    \varepsilon_{311}&\varepsilon_{312}&\varepsilon_{313}
    \end{matrix}
   \right)=
\left(\begin{matrix}
0&0&0\\
0&0&-i\\
0&i&0
\end{matrix}
\right),\\  \hat{S}_y=\hat{S}^{2}= i\left(\begin{matrix}
    \varepsilon_{121}&\varepsilon_{122}&\varepsilon_{123}\\
    \varepsilon_{221}&\varepsilon_{222}&\varepsilon_{223}\\
    \varepsilon_{321}&\varepsilon_{322}&\varepsilon_{323}
    \end{matrix}
   \right)=
\left(\begin{matrix}
    0&0&i\\
    0&0&0\\
    -i&0&0
    \end{matrix}
\right),\\  \hat{S}_z=\hat{S}^{3}= i\left(\begin{matrix}
    \varepsilon_{131}&\varepsilon_{132}&\varepsilon_{133}\\
    \varepsilon_{231}&\varepsilon_{232}&\varepsilon_{233}\\
    \varepsilon_{331}&\varepsilon_{332}&\varepsilon_{333}
    \end{matrix}
   \right)=
\left(\begin{matrix}
    0&-i&0\\
    i&0&0\\
    0&0&0
    \end{matrix}
\right).
\end{split}
\end{equation}
One can check that $[\hat{S}_x,\hat{S}_y]=i\hat{S}_z$,
$\mathbf{\hat{S}}\times\mathbf{\hat{S}}=i\mathbf{\hat{S}}$, and
$\mathbf{\hat{S}}^2=\hat{S}_x^2+\hat{S}_y^2+\hat{S}_z^2=S(S+1)$
with $S=1$. These matrices are the three generators of SU(3) group
which have eight generators called Gell-Mann matirces. Equation
(\ref{MaxwellEq}) is a relativistic Hamiltonian to discuss a
particles  with pseudospin-1, which is analogous to the Weyl
equation for the massless relativistic fermions with spin-$1/2$.

\subsection{Maxwell fermions in lattice systems}

The Maxwell Hamiltonian in Eq. (\ref{MaxwellEq})
originally describes a massless relativistic boson (photon) with
spin one. Moreover, in quantum field theory, bosons are identical
particles with zero or integer spins, while fermions are particles
with half integer spins. So it seems that the Maxwell Hamiltonian
cannot be used to describe the fermionic particles. However, there
is a fundamental difference between particles in a lattice and
those at high-energy. Rather than constrained by Poincare symmetry
in high-energy physics, quasiparticles in a lattice system are
constrained only by certain subgroups (space groups) of the
Poincare symmetry \cite{Bradlyn}. So there is the potential to
find free fermionic excitations in lattice systems for which
Hamiltonian is written in the form of Eq. (\ref{MaxwellEq}). In
the subsequent sections, we demonstrate that the Bloch Hamiltonian
of certain well-designed lattice models can be written as
\begin{equation} \label{Lattice_H}
\mathcal{H}(\mathbf{k})=\mathbf{R}(\mathbf{k})\cdot
\mathbf{\hat{S}},\end{equation}
where $\mathbf{R}(\mathbf{k})=(R_x,R_y,R_z)$ denotes the Bloch
vectors. Some threefold degenerate points exist in the bands of
the model Hamiltonian where low-energy physics should be described
by the Schr\"{o}dinger equation with the Hamiltonian
(\ref{MaxwellEq}), and thus we call such quasiparticles Maxwell
quasiparticles. Potential candidates include atoms in the optical
lattices, electrons in certain crystals, and photons in photonic
lattices. Here we focus on the fermionic atoms in optical
lattices. For these fermionic atoms, the Maxwell quasiparticles
are fermions instead of bosons (spin-1 photons) in the original
Maxwell equations. In principle, Maxwell fermions can be realized
with two different schemes. First, we can use non-interacting
fermionic atoms in a square or cubic optical lattice and choose
three atomic internal states in the ground state manifold to
encode the three spin states $|s\rangle$
($s=\uparrow,0,\downarrow$). Notably, the use of the atomic
internal degree of freedom enables us to implement our model in a
lattice of simplest geometry, i.e., a primitive square or cubic
lattice. Alternatively, Maxwell fermions can be realized by using
single-component fermionic atoms in optical lattices with three
sublattices, where the pseudospin-1 basis is replaced by the three
sublattices in a unit cell.  For conceptual simplicity, we discuss
the first scheme in the main text, and the realization of the
second scheme is addressed in \textbf{Appendix A}.

\section{Maxwell fermions in 2D lattice systems}
In this section, we construct a 2D tight-binding model on a square
lattice and then investigate the intrinsic properties of
the emergent  pseudospin-1 Maxwell fermions in different
topological phases.

\subsection{The 2D model}
The 2D model Hamiltonian we considered is given by
\begin{equation} \label{2DHam}
\begin{aligned}
\hat{H}_{2D}&=t\sum_{\mathbf{r}}\left[
\hat{H}_{\mathbf{rx}}+\hat{H}_{\mathbf{ry}}+  \left(
\Gamma_0\hat{a}^{\dag}_{\mathbf{r},0}\hat{a}_{\mathbf{r},\uparrow}
+\textrm{H.c.}\right) \right],\\
\hat{H}_{\mathbf{rx}}&=-\hat{a}^{\dag}_{\mathbf{r-x},0}(\hat{a}_{\mathbf{r},
\downarrow}+i\hat{a}_{\mathbf{r},\uparrow}) +
\hat{a}^{\dag}_{\mathbf{r+x},0}(\hat{a}_{\mathbf{r},\downarrow}-i\hat{a}_{\mathbf{r},\uparrow})+\textrm{H.c.},
\\
\hat{H}_{\mathbf{ry}}&=~~\hat{a}^{\dag}_{\mathbf{r-y},\uparrow}(\hat{a}_{\mathbf{r},\downarrow}+i\hat{a}_{\mathbf{r},0})
-
\hat{a}^{\dag}_{\mathbf{r+y},\uparrow}(\hat{a}_{\mathbf{r},\downarrow}-i\hat{a}_{\mathbf{r},0})+\textrm{H.c.},
\end{aligned}
\end{equation}
where $\hat{H}_{\mathbf{rx}}$ and $\hat{H}_{\mathbf{ry}}$ represent the spin-flip hopping along the $x$ and $y$
axis with the tunneling amplitude $t$, respectively. $\hat{a}_{\mathbf{r},s}$ is
the fermionic annihilation operator on site $\mathbf{r}$ for the
spin state $|s\rangle$, and $\Gamma_0=2iM$ with the tunable
parameter $M$ is the strength of the on-site spin-flip.

Under the periodic boundary condition, Hamiltonian
(\ref{2DHam}) can be rewritten as
$$\hat{H}_{2D}=\sum_{\mathbf{k},ss'}\hat{a}^{\dag}_{\mathbf{k}s}[\mathcal{H}(\mathbf{k})]_{ss'}\hat{a}_{\mathbf{k}s'},$$
where
$\hat{a}_{\mathbf{k}s}=1/\sqrt{V}\sum_\mathbf{r}e^{-i\mathbf{k}\cdot\mathbf{r}}\hat{a}_{\mathbf{r}s}$
is the annihilation operator in momentum space
$\mathbf{k}=(k_x,k_y)$. The Bloch Hamiltonian
$\mathcal{H}(\mathbf{k})$ has the form of Eq. (\ref{Lattice_H}),
where the Bloch vector is given by
\begin{eqnarray}
R_x &=&2t\sin{k_x},\nonumber \\
R_y &=& 2t\sin{k_y},\\
R_z &=& 2t(M-\cos{k_x}-\cos{k_y}) \nonumber
\end{eqnarray}
with the lattice spacing $a\equiv 1$ and $\hbar \equiv1$ hereafter. The
energy spectrum of this system is given by
$E(\mathbf{k})=0,\pm|\mathbf{R}(\mathbf{k})|$, which has a
zero-energy flat band in the middle of the three bands. In the following, we illustrate
that this model has rich phase diagrams: it is a normal
insulator for $|M|>2$ with Chern number $\mathcal{C}=0$; it is a
topological insulator for $|M|<2$ except of when $M=0$ with Chern
number $|\mathcal{C}|=2$; it is a topological metal for $|M|=2$ with a
quantized Berry phase $|\gamma|=2\pi$; it is a trivial metal
for $M=0$ with Berry phase $\gamma=0$ (see \textbf{Appendix B}).

\subsection{2D Maxwell fermions in Maxwell metals}
\begin{figure}[htbp]\centering
\includegraphics[width=8.7cm]{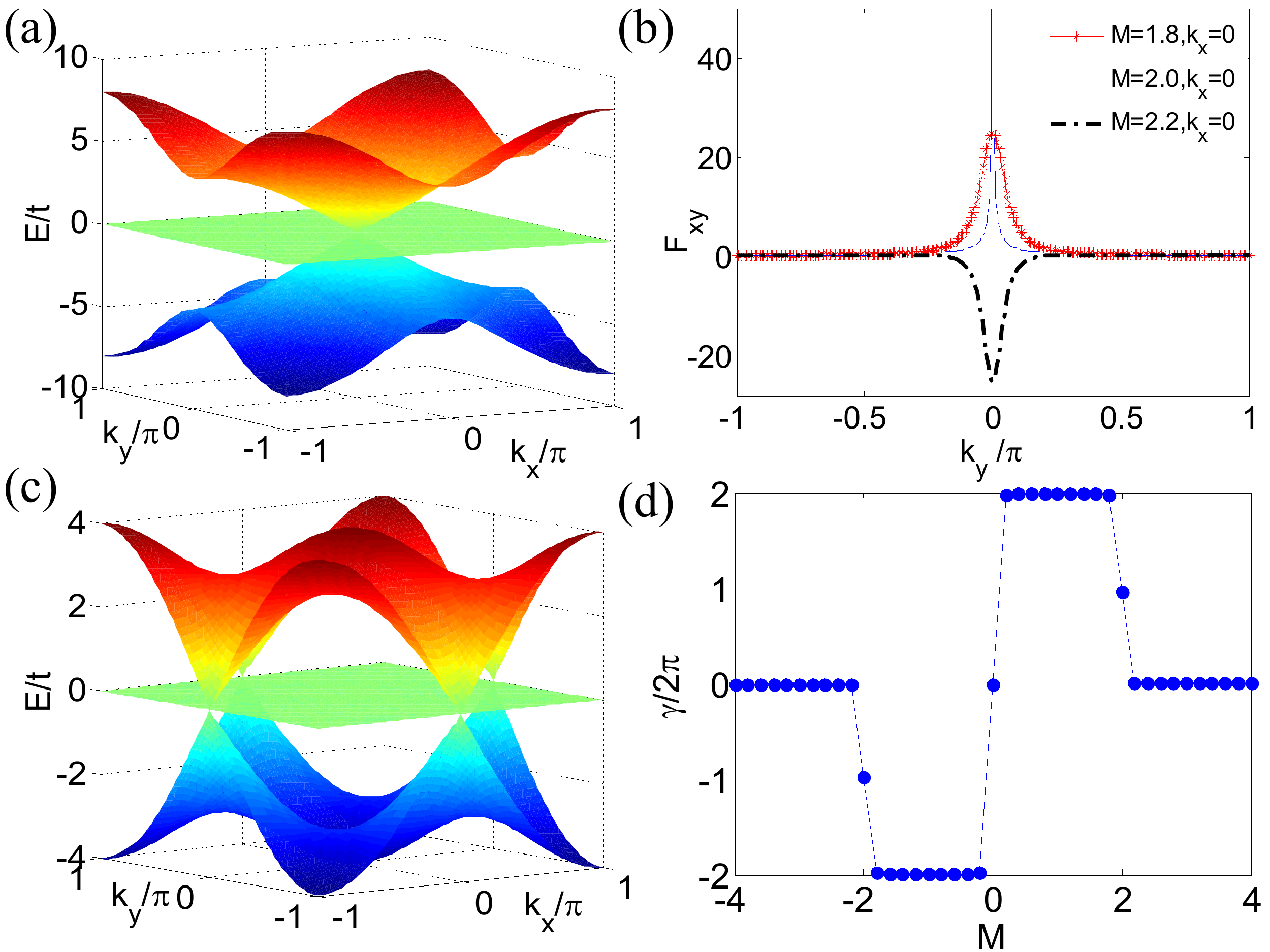}
\caption{(Color online) (a) The energy spectrum for $M=2$; (b)
The Berry curvature $F_{xy}(k_y)$ for $k_x=0$ and $M=1.8,2,2.2$;
(c) The energy spectrum for $M=0$; (d) The Berry phase $\gamma$ as
a function of the parameter $M$, which corresponds to the
 Chern number $\mathcal{C}_1=\gamma/2\pi$ when the 2D system is in the insulating phase with $M\neq0,\pm2$. \label{2D-Bulk}}
\end{figure}

The three bands touch at a single point when $M$ is $-2$ or $2$, and touch at
two points when $M$ is $0$. For $M=2$, the three bands touch at $\mathbf{K}_+=(0,0)$
in the energy spectrum shown in Fig. \ref{2D-Bulk}(a). We expand
the Bloch Hamiltonian in the vicinity of the threefold degenerate
point and obtain the following effective Hamiltonian for the
low-energy excitations in the system
\begin{equation}
\mathcal{H}_{+}(\mathbf{q})=v\emph{q}_x\hat{S}_x+v\emph{q}_y\hat{S}_y,
\end{equation}
where $v=2t$ is the effective speed of light and
$\mathbf{q}=\mathbf{k}-\mathbf{K}_+$. This effective Hamiltonian
takes the Maxwell Hamiltonian $\hat{H}_{M}$ in Eq. (\ref{MaxwellEq}) in 2D, and thus the dynamics of the low-energy excitations can
be effectively described by the Maxwell equations. In this sense, we name these low-energy
excitations Maxwell fermions and the threefold degeneracy
point  Maxwell point. When the Fermi level lies near the
Maxwell point, the system can be named Maxwell metal,
which is a metallic state due to the existence of the zero-energy
flat band.

It is interesting to note that the Maxwell point in the 2D lattice system has
topological stability characterized by a quantized Berry phase.
To study the topological stability, we calculate the Berry phase
for a Maxwell fermion circling around the Maxwell point
\begin{equation}
\gamma=\oint_cd\mathbf{k} \cdot \mathbf{\mathbf{F}(k)},
\end{equation}
where the Berry
curvature $\mathbf{F(k)}=\nabla\times \mathbf{A(k)}$ with the
Berry connection defined by the wave function
$|\psi_n(\mathbf{k})\rangle$ in the $n$-th ($n=1,2,3$) band
$\mathbf{A(k)}=-i\langle\psi_n(\mathbf{k})|\nabla_{\mathbf{k}}|\psi_n(\mathbf{k})\rangle$.
For this three-band system described by the Bloch Hamiltonian
$\mathcal{H}(\mathbf{k})$, the lowest-band Berry curvature in the
$k_x$-$k_y$ space can be rewritten as \cite{He1}
\begin{equation}
F_{xy}=-\frac{1}{R^{3}}\mathbf{R}\cdot(\partial_{k_x}{\mathbf{R}}\times\partial_{k_y}{\mathbf{R}}).
\end{equation}
The distributions $F_{xy}(k_y)$ for fixed $k_x=0$ and
typical parameters $M=1.8,2,2.2$ are plotted in Fig.
\ref{2D-Bulk}(b), and the results show that $F_{xy}$ is a
Dirac-$\delta$ function at the Maxwell point. The numerical
integration of $F_{xy}$ over the Brillouin zone for $M=2$ gives
the Berry phase $\gamma=2\pi$, which is confirmed by
analytical calculation. When $M=-2$, the single Maxwell point moves to the
Brillouin edge $\mathbf{K}_-=(\pi,\pi)$ with  a quantized Berry
phase $\gamma=-2\pi$, and the low-energy effective Hamiltonian
becomes
$\mathcal{H}_{-}(\mathbf{q})=-\mathcal{H}_{+}(\mathbf{q})$. When
$M=0$ and with the energy spectrum shown in Fig. \ref{2D-Bulk}(c),
there are two Maxwell points at $(0,\pi)$ and $(\pi,0)$ with the
effective Hamiltonian
\begin{equation}
\mathcal{H}_0(\mathbf{q})=\pm
v\emph{q}_x\hat{S}_x\mp v\emph{q}_y\hat{S}_y.
\end{equation}
In this case, the Berry phase for both Maxwell points is
$\gamma=0$, which corresponds to a trivial metallic
state. So we can conclude that the single Maxwell point with
linear dispersion relationship carrying a $\pm2\pi$ Berry phase is
a unique topological property of this 2D Maxwell metallic state,
which is different from the Dirac points in monolayer or bilayer
graphene \cite{Yao}.

\begin{figure}[htbp] \centering
\includegraphics[width=8.7cm]{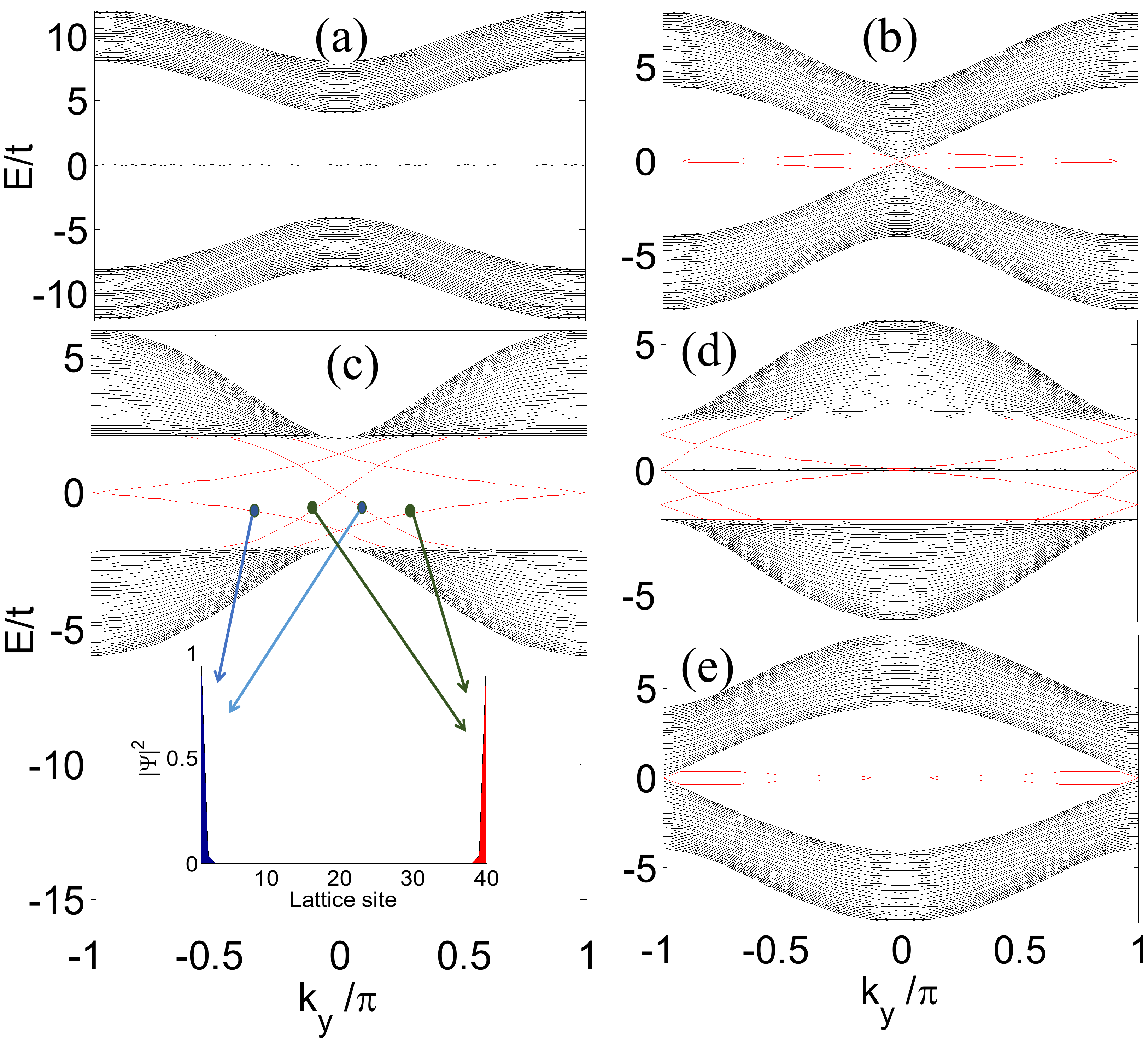}
 \caption{(Color online) Energy spectra and edge states in 2D lattices for (a) $M=4$; (b)
$M=2$; (c) $M=1$; (d) $M=-1$; and (e) $M=-2$. The inset in (c)
shows the density distributions of four typical edge modes. The
edge modes in (a-e) are plotted in red. The lattice site is $L_x=40$
under the open boundary condition.}\label{1D-Spectrum}
\end{figure}

\subsection{Maxwell edge modes in Maxwell insulators}
The system is an insulator when $M\neq0,\pm2$ since there is a gap
between any two subbands. Under this condition, we can
calculate the corresponding Chern number $\mathcal{C}_n$ for the
three bands with the band index $n$:
\begin{equation}
\mathcal{C}_n=\frac{1}{2\pi}\int_{BZ}{dk_xdk_y}F_{xy}(k_x,k_y)=\gamma/2\pi.
\end{equation}
We find that nonzero Chern numbers
$\mathcal{C}_1=-\mathcal{C}_3=2\text{sign}(M)$ for $|M|<2$ and
$\mathcal{C}_1=\mathcal{C}_3=0$ for $|M|>2$, and
thus the zero Chern number $\mathcal{C}_2(M)=0$ for the flat band (see \textbf{Appendix B}). Figure
\ref{2D-Bulk}(d) shows the Berry phase of the lowest band
$\gamma=2\pi \mathcal{C}_1$ as a function of the parameter $M$, which
indicates topological phase transition with band closing in this system when $M=-2,0,2$.

To further study the topological properties, we numerically
calculate the energy spectrum of the system under the periodic
boundary condition along the $y$ direction and under the open
boundary condition along the $x$ direction with the length $L_x =
40$. The results in Fig. \ref{1D-Spectrum} show the variation of
the energy spectra by changing the parameter $M$. For $M=4$ [Fig.
\ref{1D-Spectrum}(a)], there is no edge mode between the two band
gaps in this trivial insulating state with the Chern number
$\mathcal{C}_n=0$. When $|M|$ decreases to critical values
$M=\pm2$ [Figs. \ref{1D-Spectrum}(b) and \ref{1D-Spectrum}(e)],
the band gaps close and the system is in the nontrivial Maxwell
metallic phase with $\pm2\pi$ Berry phase (corresponding to the
Chern number $\pm1$) and a branch of edge modes connecting the
lowest (third) band and the middle flat band. For $M=\pm1$ [Figs.
\ref{1D-Spectrum}(c) and \ref{1D-Spectrum}(d)], the spectra
contain two pairs of asymmetric branches of edge modes connecting
the lowest (third) band and the middle flat band, which is
consistent with bulk-edge correspondence with the bulk Chern
number $|\mathcal{C}_{1,3}|=2$. The density distributions of some
edge modes are shown in the inset of Fig. \ref{1D-Spectrum}(c) for
typical $k_y$.

\begin{figure}[htbp]\centering
\includegraphics[width=8.7cm]{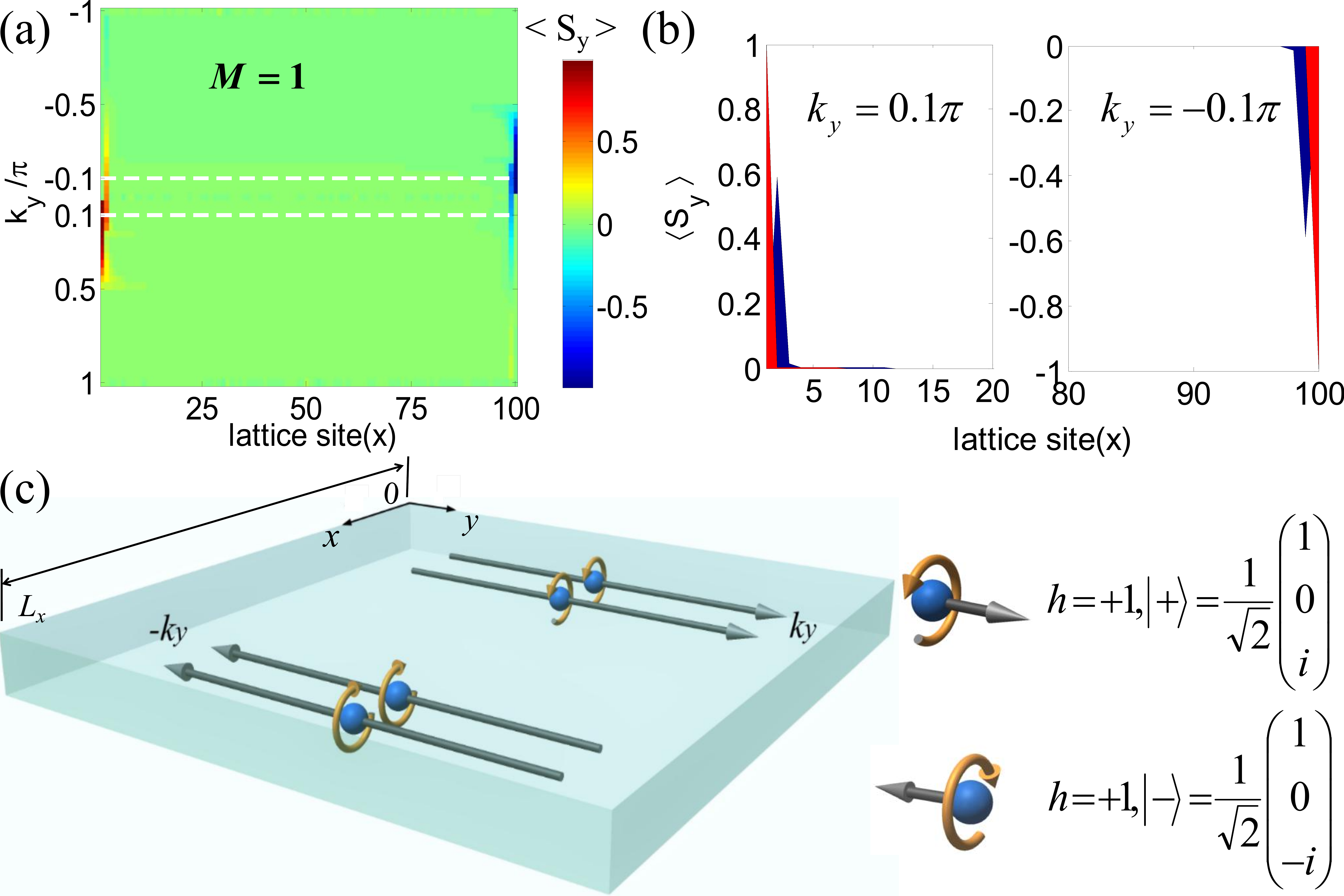}
 \caption{(Color online) (a) Expectation value of $\hat{S}_y$ as a function of $k_y$ and $x$  with lattice sites $L_x=100$ under the open boundary condition; (b) Density distribution of $\hat{S}_y(x)$ for $k_y=0.1\pi$  and $k_y=-0.1\pi$; (c) Schematic diagram for Maxwell edges states $|+\rangle$ and $|-\rangle$ in the Maxwell topological insulator with opposite momenta, both corresponding to the right circularly-polarized photons with  helicity $h=+1$.} \label{EdgeMode}
\end{figure}

The edge modes in the topological insulator phase have novel properties. Without loss of
generality, we explore the edge modes in the
first band gap for parameter $M=1$. We
find a correspondence between the helicity of these edge states
and the polarization of photons, so we named them Maxwell
edge modes in this so-called Maxwell topological insulator. In
particular, we reveal that this system exhibits the analogous
quantum anomalous Hall effect \cite{Nagaosa}, with the edge modes
being strong spin-momentum locking as eigenstates of the spin
operator $\hat{S}_y$. This means that the two bunches of
quasiparticle streams on the two edges (in the $x$ direction) can
be treated as the streams of polarized Maxwell quasiparticles
moving along the $y$ axis.

In Fig. \ref{EdgeMode}(a), we numerically calculate the expectation
value $\langle\hat{S}_y\rangle$ of the wave function with $L_x=100$.
The results show that the distribution of
$\langle\hat{S}_y(k_y,x)\rangle$ has two peaks localized at both
 the left and right edges with an opposite sign. To be more
precise, we plot $\langle\hat{S}_y(x)\rangle$ for $k_y=0.1\pi$ and
$k_y=-0.1\pi$ in Fig. \ref{EdgeMode}(b), respectively. The result
indicates that only the two edge states for each edge are the
eigenstates of $\hat{S}_y$. The edge states on the left with
positive eigenvalue are
$|+\rangle=\frac{1}{\sqrt{2}}\begin{pmatrix}1,0,i\end{pmatrix}^{T}=\frac{1}{\sqrt{2}}(\mathbf{e}_x+i\mathbf{e}_z)^T$,
and the ones on the right edge with negative eigenvalue are
$|-\rangle=\frac{1}{\sqrt{2}}\left(\begin{matrix}1,0,-i\end{matrix}\right)^{T}=\frac{1}{\sqrt{2}}(\mathbf{e}_x-i\mathbf{e}_z)^T$,
where $\mathbf{e}_j$ ($j=x,y,z$) are the unit vectors of Cartesian
coordinates. So the effective Hamiltonian of edge states is
given by
\begin{equation}
H_{\text{edge}}=v_yk_{y}\hat{S}_{y}.
\end{equation}
This effective Hamiltonian is none other than the 1D Hamiltonian of
circularly-polarized photons. The helicity operator defined as
\begin{equation}
\hat{h}=\hat{\mathbf{S}}\cdot\frac{\mathbf{k}}{|\mathbf{k}|}=\text{sign}(k_y)\hat{S}_y
\end{equation}
is the projection of the spin along the direction of the
linear momentum \cite{Lan}. Thus, the edge
quasiparticle-streams in this Maxwell topological insulator can be
treated as Maxwell fermion-streams with the same helicities
$h\equiv\langle \hat{h}\rangle=+1$ for opposite momenta, which
satisfies the helicity conservation of massless photons in quantum
field theory, as shown in Fig. \ref{EdgeMode}(c). In addition, the
momentum $\mathbf{k}$ can also be considered the wave vector of
the plane electromagnetic wave propagated along the $y$ axis.
Both edge states $|+\rangle$ $(k_y>0)$ and $|-\rangle$ $(k_y<0)$
with the same helicities can be regarded as right
circularly-polarized waves which constitute the two independent
transverse polarization vector $\mathbf{e}_x$ and $\mathbf{e}_z$
with opposite momenta. We can see from Fig. \ref{EdgeMode}(c) that
the Maxwell edge modes moving along the $+y$ ($-y$) direction
correspond to the right circularly-polarized waves rotating
anticlockwise (clockwise) in the $xz$ plane (along the $-y$
axis) propagated along the $+y$ ($-y$) direction. Likewise, when $-2<M<0$ with $\mathcal{C}=-2$, the Maxwell edge modes on the left (right)
edge with $h=-1$ correspond to the left
circularly-polarized waves propagated along the $-y$ ($+y$) direction.
Because the electromagnetic waves are transverse waves, there is
no longitudinal component and no edge mode with helicity $h=0$,
which corresponds to the unit wave vector of plane waves. So, our
Maxwell edge modes with strong spin-momentum locking
correspond perfectly to the circularly-polarized photons.

\section{Maxwell fermions in 3D lattice systems}
In this section, we generalize the proposed model and results of
Maxwell fermions to the 3D lattice system. We first construct a
3D lattice model by adding spin-flip hopping term along $z$ axis
into the previous 2D model Hamiltonian in Eq. (\ref{2DHam}), and then
study the topological properties of the Maxwell fermions near the
3D Maxwell points.

\begin{figure}[htbp]\centering
\includegraphics[width=8.7cm]{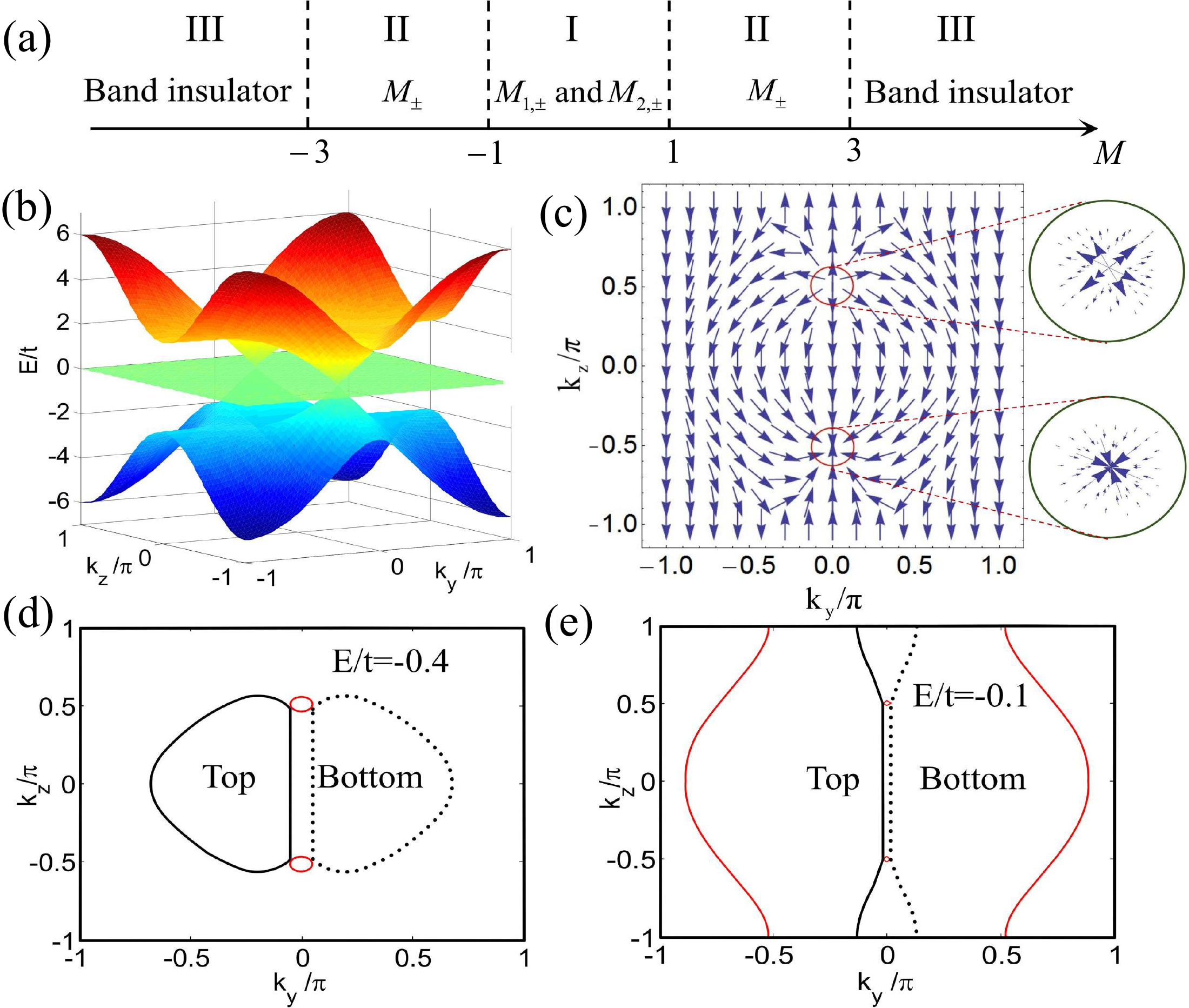}
 \caption{ (Color online) (a) The phase diagrams with regimes I, II, III
denote two pairs of Maxwell points $M_{1,\pm}$ and $M_{2,\pm}$,
one pair of Maxwell points $M_{\pm}$, and normal band insulators,
respectively. (b) The energy spectrum as a function of $k_y$ and
$k_z$ for $k_x=0$; (c) The vector distribution of the Berry
curvature $\mathbf{F}(\mathbf{k})$ at the $k_x=0$ plane. The
figures in the red  circles show $\mathbf{F(k)}$ around the
Maxwell points as a sink or source in momentum space; (d) and (e)
show Fermi arcs appear at $E/t=-0.4$ and $E/t=-0.1$ under the open
boundary condition along the $x$ direction, respectively. The two
black solid lines on the left represent two Fermi arcs connecting
the two red circles which are the bulk states near the Maxwell
points with opposite monopole charges at the top surface. The two
black dot lines on the right represent Fermi arcs in the bottom
surface. The two red lines in (e) also denote bulk states. The
parameter in (b-e) is $M=2$.} \label{phasediagram}
\end{figure}

\subsection{The 3D model}
The generalized 3D tight-binding model on a simple cubic lattice Hamiltonian is given by
\begin{equation} \label{3DHam}
\begin{split}
\hat{H}_{3D}&=~~\hat{H}_{2D}+t\sum_{\mathbf{r}}\hat{H}_{\mathbf{rz}},\\
\hat{H}_{\mathbf{rz}}&=-i(\hat{a}^{\dag}_{\mathbf{r+z},0}\hat{a}_{\mathbf{r},\uparrow}+
\hat{a}^{\dag}_{\mathbf{r},0}\hat{a}_{\mathbf{r+z},\uparrow})+\textrm{H.c.}
\end{split}
\end{equation}
where $\hat{H}_{\mathbf{rz}}$ is the additional hopping term along
the $z$ axis. The Bloch Hamiltonian of the 3D system takes the
same form as Eq. (\ref{Lattice_H}), and the Bloch vectors
$\mathbf{R}(\mathbf{k})$ with $\mathbf{k}=(k_x,k_y,k_z)$ are given
by
\begin{eqnarray}
R_x &=& 2t\sin{k_x}, \nonumber \\ R_y &=&2t\sin{k_y},\\
R_z&=&2t(M-\cos{k_x}-\cos{k_y}-\cos{k_z}). \nonumber
\end{eqnarray}
 One can check that the spin-1 matrices satisfy the following relationship under the inversion operation $\hat{P}$ and the time-reversal operation $\hat{T}$:
\begin{equation}
\begin{aligned}
\hat{P}\hat{S}_x\hat{P}^{-1}&=-\hat{S}_x,~\hat{T}\hat{S}_x\hat{T}^{-1}=-\hat{S}_x,\\
\hat{P}\hat{S}_y\hat{P}^{-1}&=-\hat{S}_y,~\hat{T}\hat{S}_y\hat{T}^{-1}=-\hat{S}_y,\\
\hat{P}\hat{S}_z\hat{P}^{-1}&=~~\hat{S}_z,~\hat{T}\hat{S}_z\hat{T}^{-1}=-\hat{S}_z,\\
\end{aligned}
\end{equation}
where $\hat{P}=\text{diag}(1,1,-1)$ and $\hat{T}=\hat{I}\hat{K}$ with
$\hat{I}=\text{diag}(1,1,1)$ and $\hat{K}$ being the complex
conjugate operator. Thus the Bloch Hamiltonian has an inversion symmetry represented by
\begin{equation}
\hat{P}\mathcal{H}(\mathbf{k})\hat{P}^{-1}=\mathcal{H}(\mathbf{-k}),
\end{equation}
but it does not have the time-reversal symmetry since
\begin{equation}
\hat{T}\mathcal{H}(\mathbf{k})\hat{T}^{-1}\neq\mathcal{H}(-\mathbf{k}).
\end{equation}

In this system, the Maxwell points can be manipulated through the tunable
parameter $M$. The phase diagram with respect to $M$ is shown in
Fig. \ref{phasediagram}(a): the system is a Maxwell metal for $|M|<3$, while it is a normal insulator for $|M|>3$. Moreover, there
are two pairs of Maxwell points denoted by
$\mathbf{M}_{1,\pm}=(0,\pi,\pm\arccos{M})$ and
$\mathbf{M}_{2,\pm}=(\pi,0,\pm\arccos{M})$ for $0\leq|M|<1$;
there are a single pair of Maxwell points at
$\mathbf{M}_{\pm}=(0,0,\pm\arccos(M-2))$
($\mathbf{M}_{\pm}=(\pi,\pi,\pm\arccos(M+2))$) for $1<M<3$
($-3<M<-1$). At the critical points of $M=\pm3$, the two Maxwell
points merge and then disappear by opening a gap when $|M|>3$,
corresponding to the normal insulating phase.

The three bands $E(\mathbf{k})=0,\pm|\mathbf{R}(\mathbf{k})|$ can
touch at certain points to form threefold degeneracy points under
the condition of $|M|\leq 3$. Considering the typical case of $M=2$,
we find that the band spectrum hosts two threefold degeneracy
points in the first Brillouin zone at
$\mathbf{M}_{\pm}=\begin{pmatrix}0,0,\pm\frac{\pi}{2}\end{pmatrix}$,
as shown in Fig. \ref{phasediagram}(b). The low-energy
effective Hamiltonian now becomes
\begin{equation} \label{MWHam}
\mathcal{H}_{M_{\pm}}(\mathbf{q})=vq_{x}\hat{S}_{x}+vq_{y}\hat{S}_{y}\pm{vq_{z}\hat{S}_{z}},
\end{equation}
where $v=2t$ is the effective speed of light. This Hamiltonian
takes the form of the isotropic 3D Maxwell Hamiltonian in Eq.
(\ref{MaxwellEq}), and thus these low-energy excitations are named 3D Maxwell
fermions.

\subsection{Topological properties}
To further study the topological properties of the 3D Maxwell fermions, we plot the
vector distribution of the Berry curvature
$\mathbf{F}(\mathbf{k})$ at the
$k_x=0$ plane in Fig. \ref{phasediagram}(c). One can find that the Maxwell points
$\mathbf{M}_{\pm}=(0,0,\pm\frac{\pi}{2})$ behave as a sink and
source of the Berry flux (see the 3D distribution of
$\mathbf{F}(\mathbf{k})$ near the two points). Thus the Maxwell
points behave like ¡°magnetic¡± monopoles in the momentum space
with topological charges defined by the Chern numbers (see \textbf{Appendix B})
\begin{equation}
\mathcal{C}_{M_{\pm}}=\frac{1}{2\pi}\oint_S d\mathbf{k} \cdot
\mathbf{\mathbf{F}(k)}=\pm2,
\end{equation}
which is twice as that of a Weyl point in Weyl
semimetals. Once we fix $k_z$ as a parameter,
$\mathcal{H}_{k_z}(k_x,k_y)$ can be viewed as a 2D Maxwell
insulator, with the $k_z$-dependent Chern number
$\mathcal{C}_{k_z}=2$ when
$k_{z}\in(-\frac{\pi}{2},\frac{\pi}{2})$ and $C_{k_z}=0$ when
otherwise. So there are always two Fermi arcs of surface states connecting a
pair of Maxwell points with opposite topological charges, as shown in Figs.
\ref{phasediagram}(d) and \ref{phasediagram}(e), in contrast
to one Fermi arc \cite{Wan,Huang,Xu1,Lv1} in Weyl semimetals. Here
the surface and bulk states are plotted in black and red, and the
surface states lie at the top and bottom surfaces along the $x$
direction are respectively denoted by solid and dot lines. It is
interesting to note that the Fermi-arc surface states in this
Maxwell metal diffuse to the bulk states due to the zero-energy
flat band, which leads to non-zero energy Fermi arcs.
In addition, the 3D Maxwell points in Maxwell metals has linear momenta along
all the three directions, in contrast to the quadratic form of
double-Weyl points \cite{Fang}.

\begin{figure} \centering
\includegraphics[width=8.6cm]{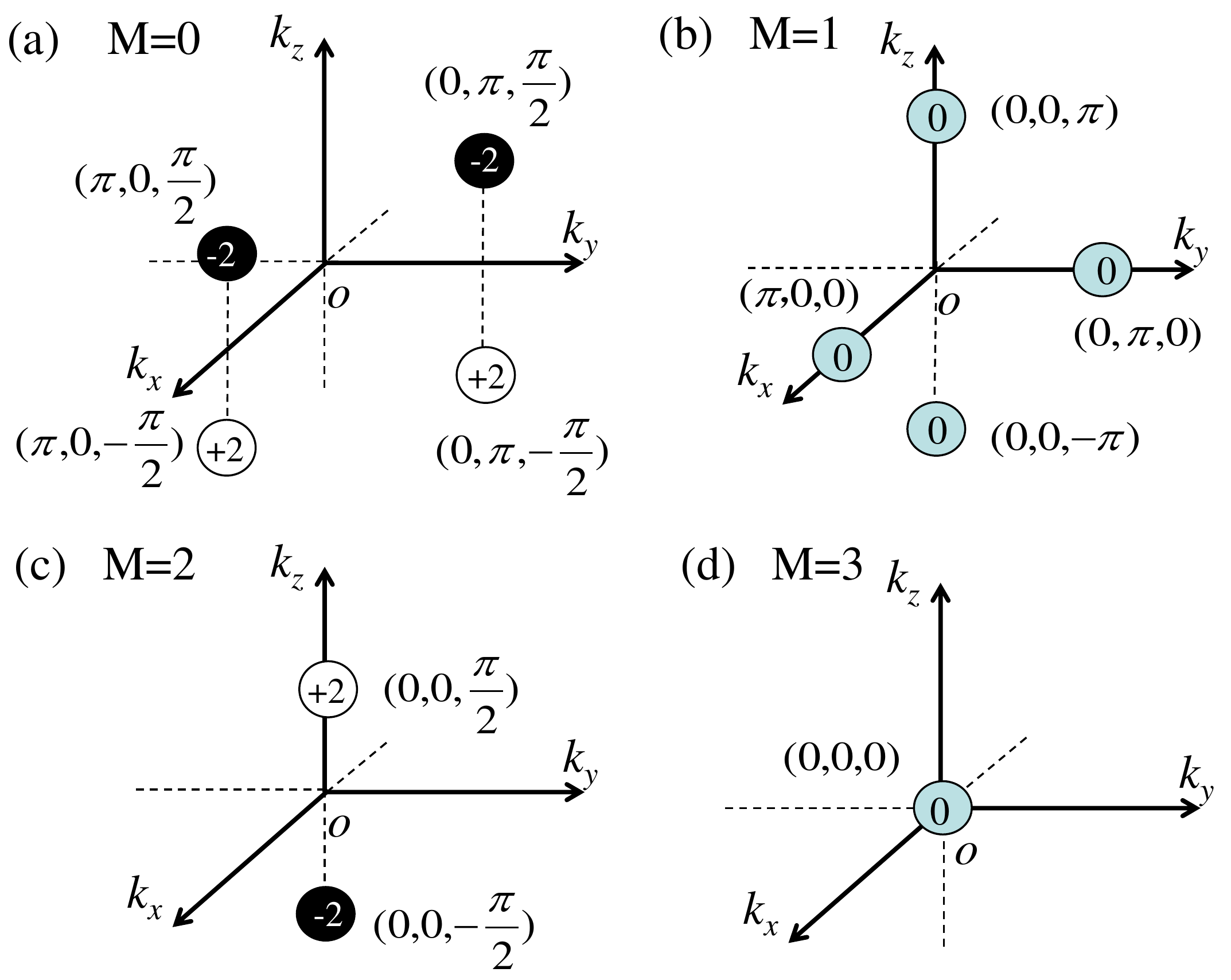}
\caption{(Color online) The merging of 3D Maxwell points. (a) When
$M=0$, two pairs of Maxwell points locate at
$(0,\pi,\pm{\frac{\pi}{2}})$ and $(\pi,0,\pm{\frac{\pi}{2}})$ with a
monopole charge of $\mp2$. (b) When $M=1$, four Maxwell points
appear at $(0,\pi,0)$, $(\pi,0,0)$ and $(0,0,\pm{\pi})$ with a zero
monopole charge. (c) Typical case of $M=2$, where there are two Maxwell
points located at $(0,0,\pm{\frac{\pi}{2}})$ with a monopole charge of
$\pm2$. (d) Critical point when $M=3$. The two Maxwell points with
opposite monopole charges emerge at $(0,0,0)$ and then disappear by
opening band gaps for $M>3$.}\label{monopole-charge}
\end{figure}

We can change the positions of the Maxwell points inside the
Brillouin zone by changing the parameter $M$. Without loss of generality, we discuss
the merging process of the Maxwell points for $M>0$. For $M=0$,
there are two pairs of Maxwell points at
$\mathbf{M}_{1,\pm}=(0,\pi,\pm{\frac{\pi}{2}})$ and
$\mathbf{M}_{2,\pm}=(\pi,0,\pm{\frac{\pi}{2}})$ with monopole
charges $\mp{2}$, as shown in Fig. \ref{monopole-charge}(a). The two pairs of Maxwell
points move together along the $k_z$ axis as we increase the
parameter $M$. They then emerge at $(0,\pi,0)$ and $(\pi,0,0)$, respectively, with a
monopole charge of $0$ for $M=1$. Meanwhile,
there are another pair of Maxwell points both with a monopole
charge of $0$ created at $(0,0,\pm{\pi})$, as shown in
Fig. \ref{monopole-charge}(b). When we continuously increase $M$, two Maxwell points
located at $(0,\pi,0)$ and $(\pi,0,0)$ disappear by opening gaps,
and the single pair of Maxwell points denoted by
$\mathbf{M}_{\pm}=(0,0,\pm{\arccos(M-2))}$ move to the $(0,0,0)$
point with a monopole charge of $\pm2$, with the typical case for
$M=2$ shown in Fig. \ref{monopole-charge}(c). At the critical point $M=3$ in Fig.
\ref{monopole-charge}(d), the two Maxwell points with opposite monopole charges emerge
at $(0,0,0)$ and then disappear by opening gaps when $M>3$.

Finally in this section, we brief discuss the
topological stability of the 3D Maxwell points. Unlike Weyl points
that are  strongly robust \cite{Wehling}, Maxwell points can only
be stabilized by certain symmetry. In our proposed lattice
systems, the band gaps will be opened and thus Maxwell points are
gapped when the inversion symmetry is broken by introducing a
perturbation term with one of the other five Gell-Mann matrices.
This is due to the fact that the Maxwell Hamiltonian only takes
three of the eight Gell-Mann matrices. So the topological
stability of Maxwell points are weaker than Weyl points. However,
Maxwell points would still be stable as long as the system is
protected by certain symmetry, such as the inversion symmetry in
our 3D system. In this case, the perturbations under the same
symmetry would not open the band gaps and only change the
positions of Maxwell points in the Brillouin Zone (see Fig.
\ref{monopole-charge}) until the topological phase transition
occurs.

\section{Proposal for experimental realization and detection}
Now we proceed to propose a realistic scheme for realizing the 2D
and 3D model Hamiltonians with ultracold atoms in the square and
cubic  optical lattices by using the Raman-assisted tunneling
method \cite{LAT1,LAT2,LAT3,LAT4,LAT5,LAT6,LAT7}, respectively. We
also suggest that the Chern numbers can be revealed from the shift
of the hybrid Wannier center of an atomic cloud, based on a
generalization of topological pumping in the optical lattices
\cite{Thouless,Marzari,Smith,Wang,Pumping1,Pumping2,Pumping3}.

\begin{figure*}[htbp]
\centering
\includegraphics[width=14cm]{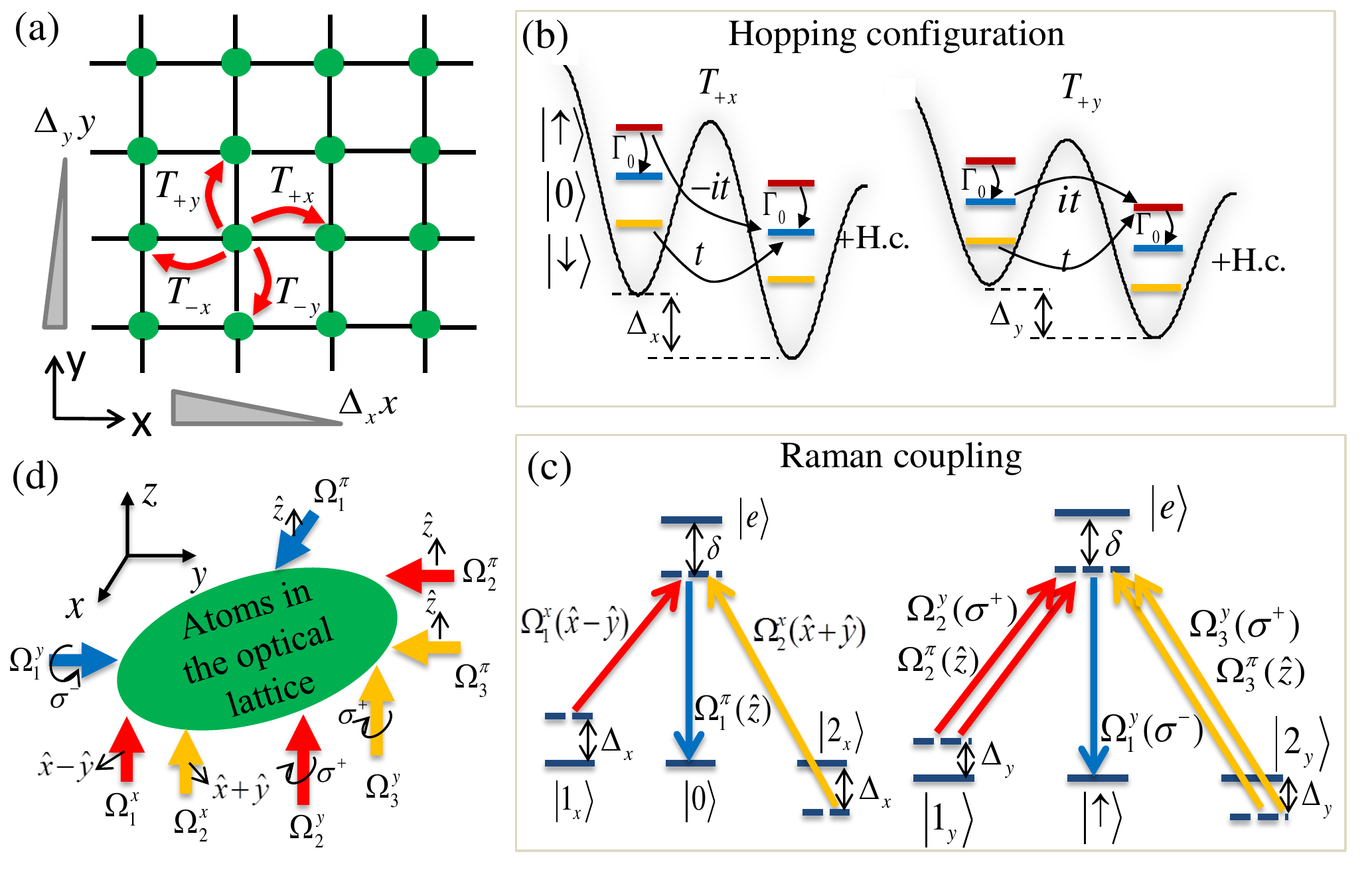}
\caption{Scheme for realizing the 2D model Hamiltonian. (a)
Schematic diagram of the square optical lattice and the atomic
hopping along the two  axis denoted by $T_{\pm x}$ and $T_{\pm
y}$. (b) The atomic hopping configuration in the 2D model
Hamiltonian in Eq. (\ref{2DHam}). The three atomic internal states
$|\uparrow\rangle,|0\rangle,|\downarrow\rangle$ form the (pseudo)spin-1 basis, such that $T_{\pm (x,y)}$
represent spin-flip hopping and the $\Gamma_0$ term represent
on-site spin-flip.  A large linear tilt $\Delta_{x,y}$ per lattice
site along $x,y$-direction is used to suppress the natural hopping
and the hopping can then be restored by Raman lasers. (c) The
Raman coupling scheme for engineering the required spin-flip
hopping along each axis. The detuning in each direction matches
the frequency offset of the corresponding Raman beams. (d) The
Raman lasers with the corresponding polarization and propagation
direction.} \label{2dOL}
\end{figure*}

\subsection{Proposed realization in optical lattices}

To realize the 2D model Hamiltonian in Eq. (\ref{2DHam}), we consider noninteracting atoms in a titled square optical lattice with the lattice spacing $a$ and choose three atomic internal states in the ground state manifold to carry the spin states, as shown Fig. \ref{2dOL}(a). The other levels in the ground state manifold are irrelevant as they can be depopulated by the optical pumping. The on-site spin-flip term $\Gamma_0\hat{a}^{\dag}_{\mathbf{r},0}\hat{a}_{\mathbf{r},\uparrow}$ can be easily achieved by applying a radio-frequency field or Raman beams for coupling the atomic internal states $|\uparrow\rangle$ and $|0\rangle$. The major difficulty for implementing our model is to realize the spin-flip hopping terms $\hat{H}_{\mathbf{rx}}$ and $\hat{H}_{\mathbf{ry}}$ along each direction. By defining the superposition states
\begin{equation}
\begin{aligned}
|1_x\rangle&=\left(|\downarrow\rangle-i|\uparrow\rangle\right)/\sqrt{2},~ |2_x\rangle=\left(|\downarrow\rangle+i|\uparrow\rangle\right)/\sqrt{2},\\
|1_y\rangle&=\left(|\downarrow\rangle-i|0\rangle\right)/\sqrt{2},~~
|2_y\rangle=\left(|\downarrow\rangle+i|0\rangle\right)/\sqrt{2},
\end{aligned}
\end{equation}
the two hopping terms can be diagrammatically visualized as
\begin{equation}
 \begin{aligned}
T_x&=T_{+x}+T_{-x}\\
&=\overset{\times }{
\curvearrowleft }|1_x\rangle\overset{\sqrt{2}}{\curvearrowright }|0\rangle+|0\rangle\overset{-\sqrt{2}}{\curvearrowleft}%
|2_x\rangle\overset{\times }{\curvearrowright }+\text{
H.c.},\\
T_y&=T_{+y}+T_{-y}\\
&=\overset{\times }{
\curvearrowleft }|1_y\rangle\overset{-\sqrt{2}}{\curvearrowright }|\uparrow\rangle+|\uparrow\rangle\overset{\sqrt{2}}{\curvearrowleft}%
|2_y\rangle\overset{\times }{\curvearrowright }+\text{
H.c.},
\end{aligned}
\end{equation}
where $\overset{\times }{\curvearrowright }$ indicates that the hopping is
forbidden along this direction. The hopping terms $T_{+x}$ and $T_{+y}$ are shown in Fig. \ref{2dOL}(b), which can be realized by using the Raman-assisted tunneling with proper laser-frequency and polarization selections \cite{LAT1,LAT2,LAT3,LAT4,LAT5,LAT6,LAT7}. First, the required broken parity (left-right) symmetry is achieved by titling the lattice with a homogeneous energy gradient along the two directions, which can be created by the gravitational field or the gradient of a dc- or ac-Stark shift. Here we require different linear energy shifts per site $\Delta_{x,y}$ along different directions, such as $\Delta_x=1.5\Delta_y$. Then the natural hopping is suppressed by the large tilt potential, and the hopping terms are restored and engineered by applying two-photon Raman coupling with the laser beams of proper configurations [see Fig. \ref{2dOL}(c)].

We consider the realization of the hopping term $\hat{H}_{\mathbf{rx}}$ to
explain the Raman scheme, and first focus on the single term $T_{+x}^{(1)}=\hat{a}^{\dag}_{\mathbf{r+x},0}(\hat{a}_{\mathbf{r},\downarrow}-i\hat{a}_{\mathbf{r},\uparrow})$ (here $T_{+x}=T_{+x}^{(1)}+\text{
H.c.}$) for details. This term corresponds to an atom in the spin state $|1_x\rangle=\left(|\downarrow\rangle-i|\uparrow\rangle\right)/\sqrt{2}$ at site $\mathbf{r}$ hopping to site $\mathbf{r+x}$ while changing the spin state to $|0\rangle$ with hopping strength $\sqrt{2}t$, which can be visualized as
$\overset{\times }{
\curvearrowleft }|1_x\rangle\overset{\sqrt{2}}{\curvearrowright }|0\rangle$.
This hopping term can be achieved by two Raman beams $\Omega _{1}^{x}(\hat{x}-\hat{y})=\sqrt{2}\Omega _{0}e^{ikz}$ polarized along $(\hat{x}-\hat{y})$-direction and $\Omega _{1}^{\pi}(\hat{z})=\Omega _{0}e^{ikx}$ with $\pi$-polarization along $\hat{z}$-direction. Here the population of the excited state $|e\rangle$ which is estimated by $|\Omega_{0}/\delta|^{2}$ is negligible due to the large single-photon detuning $\delta$. The two-photon detuning $\Delta_{x}$ matching the linear energy shift of the lattice per site ensures that it only allows $|1_x\rangle$ hopping to the right, and the other direction is forbidden by a large energy mismatch $2\Delta_{x}$. We can address the spin states through the polarization selection rule since the original spin basis $|\downarrow\rangle,|0\rangle, |\uparrow\rangle$ differ in the magnetic quantum number by one successively. Thus, a $\pi$-polarized beam $\Omega_{1}^{\pi}$ excites the state $|0\rangle$ and a linear $(\hat{x}-\hat{y})$-polarized beam $\Omega_{1}^{x}$ excites the superposition state $|1_x\rangle=\left(|\downarrow\rangle-i|\uparrow\rangle\right)/\sqrt{2}$ as the polarization $(\hat{x}-\hat{y}) \sim (\sigma^{+}-i\sigma^{-})$. These two beams together induce a Raman-assisted hopping between $|1_x\rangle$ and $|0\rangle$. The hopping amplitude and phase are controlled by the corresponding Raman beam amplitude and phase \cite{LAT1,LAT2,LAT3,LAT4,LAT5,LAT6,LAT7}, which can be written as
\begin{eqnarray}
t_{\mathbf{r,+x}}&=&\frac{\sqrt{2}|\Omega_0|^2}{\delta}\beta
e^{i\delta \mathbf{k}\cdot\mathbf{r}}, \\ \nonumber
\beta&=&\int dxw^{\ast
}(x+a)e^{-ikx}w(x)\int dyw^{\ast }(y)w(y).
\end{eqnarray}
Here $\delta \mathbf{k}=(-k,0)$ and we have used factorization of the Wannier function
$w(\mathbf{r}^{\prime})=w(x^{\prime })w(y^{\prime })$ in the square lattice. If we adjust the interfering
angle of the lattice beams to satisfy the condition $ka=2\pi$, the
site dependent phase term can always be reduced to $e^{i\delta
\mathbf{k}\cdot \mathbf{r}}=1$. In this case, we can obtain the
required hopping strength $t_x=\sqrt{2}t$ with $t=\beta|\Omega_0|^2/\delta$. We note that these two Raman lasers simultaneously induce the hermitian conjugate process of the hopping $T_{+x}^{(1)}$, which is the desired spin-flip tunneling from $|0\rangle$ to $|1_x\rangle$ along the $+x$ direction. Similarly, the second hopping term
$T_{-x}$ can be realized by the two Raman beams $\Omega
_{1}^{\pi}(\hat{z})=\Omega _{0}e^{ikx}$ and $\Omega
_{2}^{x}(\hat{x}+\hat{y})=-\sqrt{2}\Omega _{0}e^{ikz}$ polarized
along $(\hat{x}+\hat{y})$-direction, which couple the state
$|0\rangle$ and $|2_x\rangle$ since $(\hat{x}+\hat{y}) \sim
(\sigma^{+}+i\sigma^{-})$. Thus the hopping along the $x$ axis can thus
be realized by three Raman beams with the configuration shown in Fig. \ref{2dOL}(c).

The hopping terms $\hat{H}_{\mathbf{ry}}$ along the $y$ axis can be
realized in the similar way. Along this direction, the hopping term $T_{+y}=- \hat{a}^{\dag}_{\mathbf{r+y},\uparrow}(\hat{a}_{\mathbf{r},\downarrow}-i\hat{a}_{\mathbf{r},0})+\text{H.c.}$ can be realized by three Raman beams $\Omega _{1}^{y}(\sigma^-)=\Omega _{0}e^{iky}$ which excites the state $|\uparrow\rangle$, and $\Omega _{2}^{y}(\sigma^+)=-\sqrt{2}\Omega _{0}e^{ikz}$ and $\Omega _{2}^{\pi}(\hat{z})=i\sqrt{2}\Omega _{0}e^{-iky}$ which together effectively excite the state $|1_y\rangle=\left(|\downarrow\rangle-i|0\rangle\right)/\sqrt{2}$. The hopping term $T_{-y}=\hat{a}^{\dag}_{\mathbf{r-y},\uparrow}(\hat{a}_{\mathbf{r},\downarrow}+i\hat{a}_{\mathbf{r},0})+\text{H.c.}$ can be realized by two additional Raman beams $\Omega _{3}^{y}(\sigma^+)=\sqrt{2}\Omega _{0}e^{ikz}$ and $\Omega _{3}^{\pi}(\hat{z})=i\sqrt{2}\Omega _{0}e^{-iky}$ which effectively excite the state $|2_y\rangle=\left(|\downarrow\rangle+i|0\rangle\right)/\sqrt{2}$. Here a wave-vector difference $\delta \mathbf{k}' = (0,-2k)$ leads to the same hopping strength $t_y=t_x$ along this axis. The laser configurations for realizing the desired hopping terms along the $y$ axis are also shown in Fig. \ref{2dOL}(c). The total laser beams with the polarization and propagation directions for realizing all the hopping terms in the 2D model Hamiltonian in Eq. (\ref{2DHam}) are shown in Fig. \ref{2dOL}(d). Note that the detuning in each direction matches the frequency offset of the corresponding Raman beams. It is also important to forbid the undesired tunneling terms that requires different linear energy shifts per site along the two axis, which can be achieved by adjusting the direction of the gradient field to be in different angles with respect to the axes of the square optical lattice.

\begin{figure}[htbp]
\centering
\includegraphics[width=8.5cm]{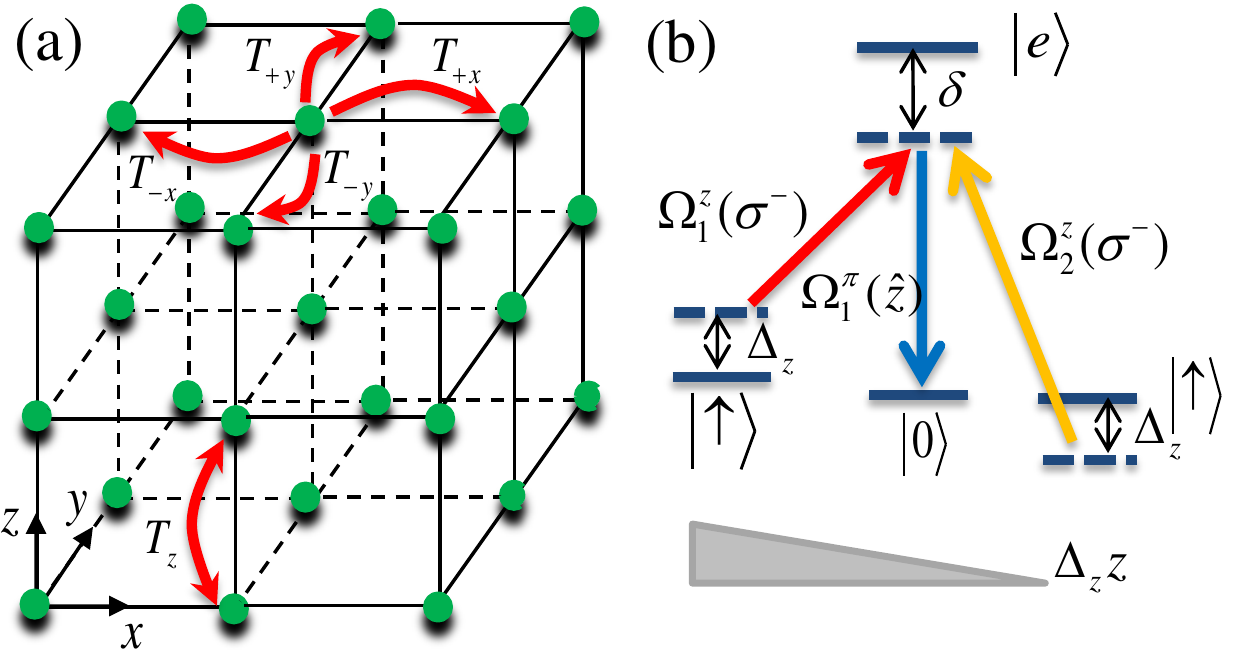}
 \caption{(a) Schematic diagram of the simple cubic optical lattice and the atomic hopping. (b) The Raman coupling for realizing the hopping term $T_z$ along the $z$ axis. } \label{3dOL}
\end{figure}

The proposed Raman scheme can be directly extended to realize the 3D model Hamiltonian in Eq. (\ref{3DHam}). In this case, one can prepare the noninteracting atoms in a titled cubic optical lattice, as shown in Fig. \ref{3dOL}(a). Here we require different linear energy shifts per site $\Delta_{x,y,z}$ along different directions, such as $\Delta_x=1.5\Delta_y=2\Delta_z$. The additional hopping $\hat{H}_{\mathbf{rz}}$ along the $z$ axis can be
implemented with the similar Raman coupling. Along the $z$ axis, the hopping term can be diagrammatically visualized as
\begin{equation}
\begin{aligned}
T_z& = \overset{\times }{
\curvearrowleft }|\uparrow\rangle\overset{-i}{\curvearrowright }|0\rangle+|0\rangle\overset{-i}{\curvearrowleft}%
|\uparrow\rangle\overset{\times }{\curvearrowright }+\text{
H.c.}.
\end{aligned}
\end{equation}
Combing with the laser $\Omega_1^{\pi}$ that couples the state $|0\rangle$, two additional Raman beams $\Omega_1^z(\sigma^{-})=-i\Omega_0e^{ikz}$ and $\Omega_2^z(\sigma^{-})=-i\Omega_0e^{-ikz}$ that couple the state $|\uparrow\rangle$ can be used to induce the hopping $\overset{\times }{\curvearrowleft }|\uparrow\rangle\overset{-i}{\curvearrowright }|0\rangle$ and $|0\rangle\overset{-i}{\curvearrowleft}%
|\uparrow\rangle\overset{\times }{\curvearrowright }$, respectively. The laser configurations for realizing the desired hopping term along the $z$ axis are shown in Fig. \ref{3dOL}(b), and thus all the hopping terms along each direction in the 3D model Hamiltonian can be realized.

Although the implementation of the 2D and 3D model Hamiltonians involves a number of Raman beams, all of the lasers can be drawn from a same one with the small relative frequency shift induced by an acoustic optical modulator. In addition, one can lock the relative frequency differences of these beams by the driving fields of the modulator such that the absolute frequencies and their fluctuations are not important. In typical experiments, for instance, by using $^{40}$K atoms of mass $m$ in an optical lattice with the lattice constant $a=2\pi
/k=764$ nm, the gravity induces a potential gradient per lattice site $\Delta =mga/\hbar\approx 2\pi \times 0.75\,$kHz. Gravity can provide the required gradients along three directions with an appropriate choice of the relative axes of the frame to satisfy $\Delta _{z}:\Delta _{y}:\Delta _{x}=1:1.5:2$ and $\Delta=\sqrt{\Delta _{x}^{2}+\Delta _{y}^{2}+\Delta _{z}^{2}}$, such that
$\Delta _{z}\approx 2\pi \times 1\,$kHz. For a lattice with potential depth $V_{0}\approx 20E_{r}$,
where $E_{r}=\hbar ^{2}k^{2}/2m$ is the recoil energy, the overlap ratio $%
\beta \approx 0.01$ and the natural tunneling rate $t_{N}/\hbar \sim 10^{-3}E_r/\hbar\approx 2\pi \times 8.5\,$Hz. For Raman beams with $\Omega _{0}/2\pi \approx 130\,$MHz and the single-photon detuning $\delta /2\pi \approx 1.7\,$THz, one has $\left\vert \Omega _{0}\right\vert ^{2}/\delta \approx 2\pi \times 10\,$kHz and the Raman-assisted hopping rate $t/\hbar \approx 2\pi \times 0.1\,$kHz. Thus the population of the excited state $|e\rangle$ which is estimated by $|\Omega_{0}/\delta|^{2}$ is negligible due to the large single-photon detuning $\delta$. During the typical
experimental time of the order of $10/t$, the undesired off-resonant hopping probabilities with upper bounded by $t_{N}^{2}/\Delta_{x}^{2}\sim0.01$ and the effective spontaneous emission rate estimated by $|\Omega _{0}/\delta |^{2}\Gamma _{s}$ with the decay rate of the excited state $\Gamma _{s}\approx 2\pi \times 6\,$MHz would be negligible \cite{LAT6}. We note that using several Raman lasers will lead to considerable heating effects in realistic experiments \cite{LAT7}, which is the main disadvantage in our proposal. The cold atom system can be effectively described by the proposed model Hamiltonian as long as the duration of an experiment is short compared to the heating time (the typical lifetime of an atomic gas), which is about 100 ms in experiments containing the heating induced by the Raman couplings \cite{LAT4,LAT7}.

\subsection{Proposed detections}

The 2D and 3D Maxwell points in the band structures that have related topological phase transition can be
detected by the Bragg spectroscopy or Bloch-Zener oscillations,
similar to the methods used for detecting Dirac and Weyl points in optical
lattices \cite{Zhu,Tarruell,Zhang2012,Zhang2015}. In addition, the Berry curvature and
thus the quantized Berry phases and the Chern numbers can be measured
by the newly-developed technique of tomography of Bloch bands in
optical lattices \cite{BandTomography,Li}, and the Chern numbers
can also be revealed from the shift of an atomic cloud's center-of-mass
\cite{Bloch2015}. Below we propose that the Chern numbers can also be revealed
from the shift of the hybrid Wannier center of an atomic cloud, based on a generalization of topological pumping in optical lattices \cite{Thouless,Marzari,Smith,Wang,Pumping1,Pumping2,Pumping3}.

The 2D Maxwell insulator can be viewed as a fictitious one-dimensional insulator subjected to an external parameter $k_y$ by using the dimension reduction method \cite{Wang}. Thus, its Chern number can be defined by the polarization
\begin{equation}
P=\frac{1}{2\pi}\int_{-\pi}^{\pi} \mathbf{A(k)}dk_x
\end{equation}
for the geometry of the underlying band structure. According to the modern theory of polarization, the Chern number defined in $k_x$-$k_y$ space can be obtained from the change in polarization induced by adiabatically changing the parameter $k_y$ by $2\pi$:
\begin{equation}
\mathcal{C}=\int_{-\pi}^{\pi}\frac{\partial P(k_y)}{\partial k_y}dk_y.
\end{equation}
For measuring $P(k_y)$, one can use another fact that the polarization can alternatively written as the center of mass of the Wannier function constructed for the single occupied band. In this system, the polarization $P(k_y)$ can be expressed by means of the centers of the hybrid Wannier functions, which are localized in the $x$ axis retaining Bloch character in the $k_y$ dimension. The variation of the polarization and thus the Chern number are directly related to the shift of the hybrid Wannier center along the $x$ axis in the lattice. The shift of hybrid Wannier center by adiabatically changing $k_y$ is proportional to the Chern number, which is a manifestation of topological pumping with $k_y$ being the adiabatic pumping parameter

In the 2D lattice system, to construct and calculate the hybrid Wannier center, we can consider the Bloch Hamiltonian with parameter $k_y$ and recover it to the tight-binding Hamiltonian along the $x$ axis. The hybrid Wannier center can be written as \cite{Wang}
\begin{eqnarray}
\langle n_x(k_y) \rangle = \frac{\sum_{i_x}i_x\rho(i_x,k_y)}{\sum_{i_x}\rho(i_x,k_y)},
\end{eqnarray}
where $\rho(i_x,k_y)$ is the density distribution of hybrid Wannier center and denotes the atomic densities resolved along the $x$ direction as a function of $k_y$. The density distribution can be written as
\begin{eqnarray}
\rho(i_x,k_y)=\sum_{\text{occupied states}}\langle i_x,k_y|i_x,k_y\rangle,
\end{eqnarray}
where $|i_x,k_y\rangle$ is the hybrid eigenstate of the system. In cold atom experiments, the atomic density distribution $\rho(i_x,k_y)$ can be directly measured by the hybrid time-of-flight images, which is referring to an \textit{in situ} measurement of the density distribution of the atomic cloud in the $x$ direction during free expansion along the $y$ direction. In the measurement, the optical lattice is switched off along the $y$ and $z$ directions while keeping the system unchanged in the $x$ direction. One can map out the crystal momentum distribution along $k_y$ in the time-of-flight images and a real space density resolution in the $x$ direction can be done at the same time. Thus one can directly extract Chern number from this hybrid time-of-fight images in the cold atom system.

\begin{figure}[htbp] \centering
\includegraphics[width=8.6cm]{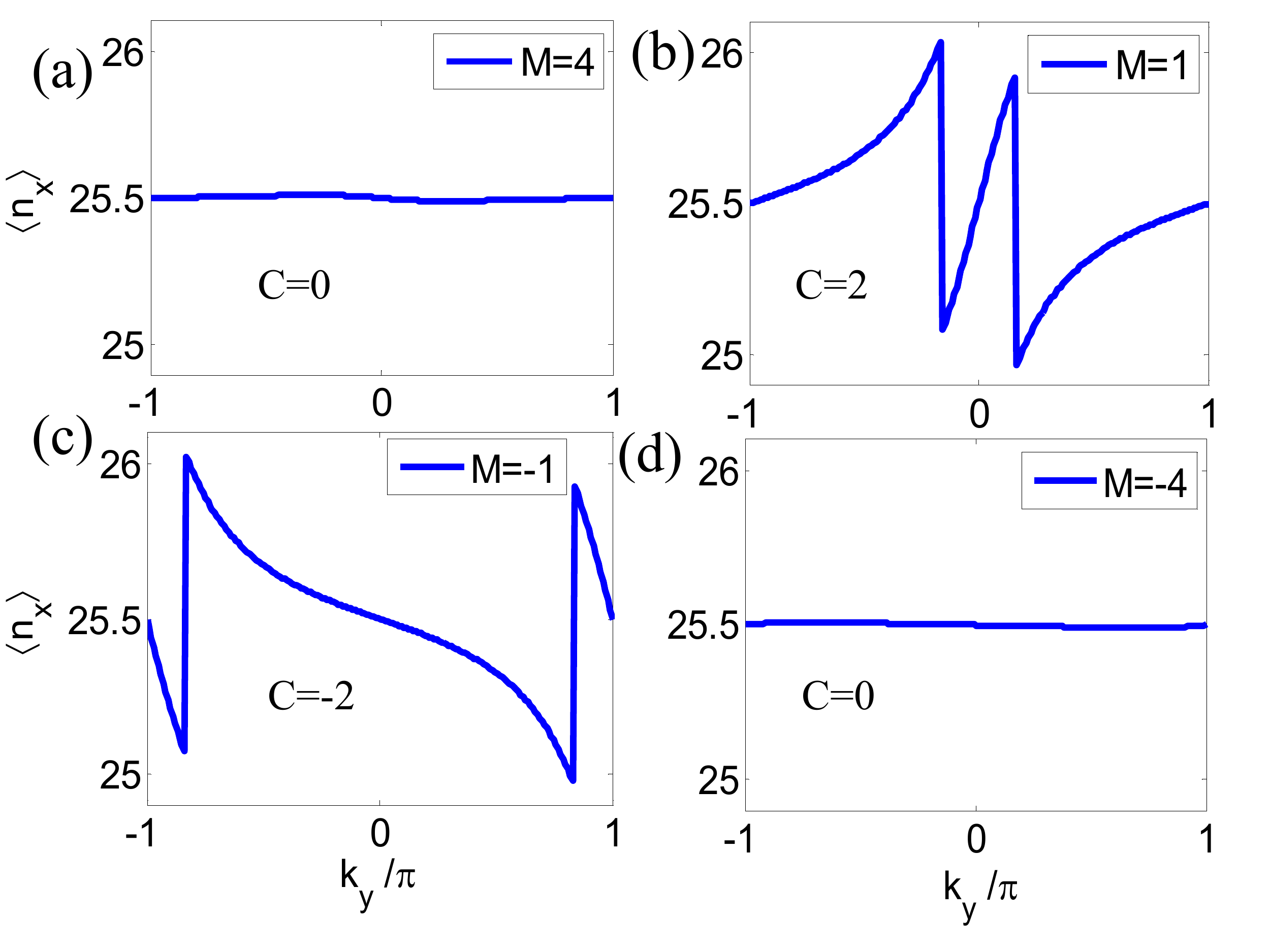}
 \caption{(Color online) The hybrid Wannier centers in a tight-binding chain of length $L_x=50$ at $1/3$ filling as a function of the adiabatic pumping parameter $k_y$ for different parameters $M$. (a) and (d) without jump of $\langle{n_{x}(k_y)}\rangle$, which is consistent with the expected Chern number of the lowest band $\mathcal{C}=0$ for $M>2$ and $M<-2$ in the 2D trivial insulator phase. In (b) and (c), $\langle{n_{x}(k_y)}\rangle$ both show the jump of two unit cell for $M=1$ and $M=-1$, corresponding to nontrivial Maxwell insulator phase with $\mathcal{C}=2$ and $\mathcal{C}=-2$, respectively. } \label{2DHWfs}
\end{figure}

We perform numerical simulations to demonstrate the feasibility of the detection scheme. We numerically calculate $\langle{n_{x}(k_y)}\rangle$ in a tight-binding chain of length $L_x=50$ at $1/3$ filling (assuming the Fermi energy $E_F=0$), and the results for typical parameters are shown in Fig. \ref{2DHWfs}. For the case in Figs. \ref{2DHWfs}(a) and \ref{2DHWfs}(d) when $M=4$ and $M=-4$, respectively, $\langle{n_{x}(k_y)}\rangle$ shows no jump, which are consistent with the expected Chern number of the lowest band $\mathcal{C}=0$ for the trivial cases. The results for $M=1$ in Fig. \ref{2DHWfs}(b) shows two discontinuous jumps of one unit cell, indicating that a particle is pumped across the system \cite{Marzari,Smith}, the Chern number of the lowest band for this case is $\mathcal{C}=2$, and the result in Fig. \ref{2DHWfs}(c) is similar to Fig. \ref{2DHWfs}(b) but with the opposite jump direction, indicating $\mathcal{C}=-2$ when $M=-1$. This establishes a direct and clear connection between the shift of the hybrid density center and the topological invariant.

\begin{figure}[htbp] \centering
\includegraphics[width=8.5cm]{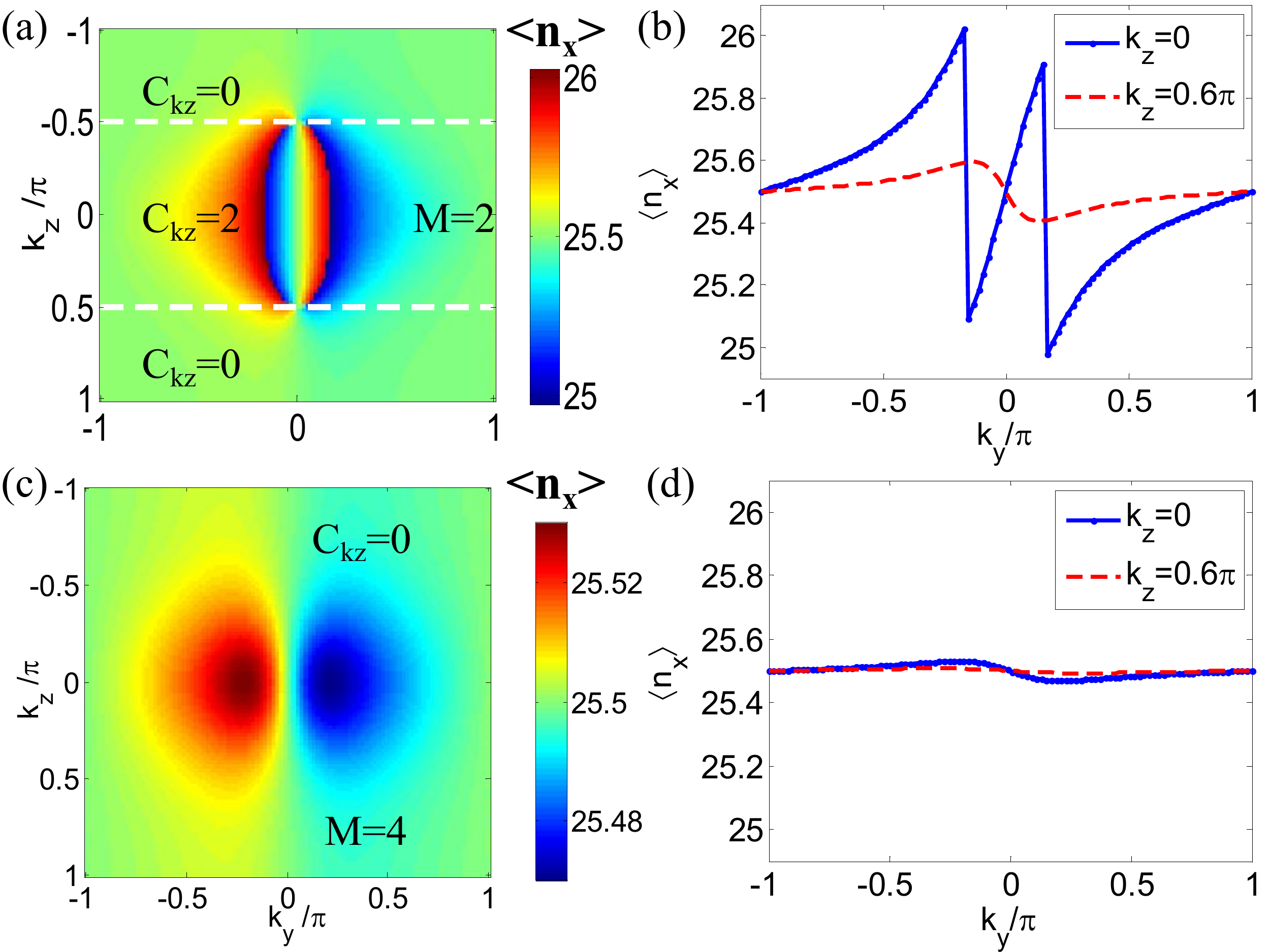}
\caption{ The hybrid Wannier centers in a tight-binding chain of length $L_x=50$ at $1/3$ filling as a function of the adiabatic pumping parameter $k_y$ for different parameters $M$ and $k_z$. (a) The profile $\langle n_x(k_y,k_z)\rangle$ for the parameter $M=2$, where $\langle n_x(k_y)\rangle$ (do not) shows double jumps in one unit cell for $k_z$ (outside) within the region $(-\frac{\pi}{2},\frac{\pi}{2})$, with typical examples shown in (b). (c) The profile $\langle n_x(k_y,k_z)\rangle$ for $M=4$ shows no jump of $\langle n_x(k_y)\rangle$ for the whole $k_z$ regime, with typical examples shown in (d). The corresponding $k_z$-dependent Chern number $\mathcal{C}_{k_z}$ is also plotted both in (a) with the white dashed lines denoting the critical value between the trivial regimes with $\mathcal{C}_{k_z}=0$ and nontrivial regimes with $\mathcal{C}_{k_z}=2$.} \label{3DHWfs}
\end{figure}

This method can be extended to detect the 3D Maxwell system. The 3D system can be further treated as a collection of $k_z$-modified 2D trivial or nontrivial Maxwell insulators with the $k_z$-dependent Chern number $\mathcal{C}_{k_z}$ defined in the $k_x$-$k_y$ plane as different slices of out-of-plane quasimomentum $k_z$:
\begin{equation}
\mathcal{C}_{k_z}=\int_{-\pi}^{\pi}\frac{\partial P(k_y,k_z)}{\partial k_y}dk_y,
\end{equation},
with the modified polarization $P(k_y,k_z)$ for the reduced one-dimensional insulators. In this case, the corresponding hybrid Wannier center is given by
\begin{eqnarray}
\langle n_x(k_y,k_z) \rangle = \frac{\sum_{i_x}i_x\rho(i_x,k_y,k_z)}{\sum_{i_x}\rho(i_x,k_y,k_z)},
\end{eqnarray}
which can also be measured by the hybrid time-of-flight images. The typical numerical the results of $\langle n_x(k_y,k_z)\rangle$ in a tight-binding chain of length $L_x=50$ at $1/3$ filling are shown in Fig. \ref{3DHWfs}. For $M=2$ in Figs. \ref{3DHWfs}(a) and \ref{3DHWfs}(b), the two 3D Maxwell points at $k_z=\pm\frac{\pi}{2}$ separate the band insulators with $\mathcal{C}_{k_z}=0$ and the topological insulators with $\mathcal{C}_{k_z}=2$. As shown in Fig. \ref{3DHWfs}(a), the hybrid Wannier center $\langle n_x(k_y)\rangle$ exhibits two discontinuous jumps of one unit cell within the region $k_z\in(-\frac{\pi}{2},\frac{\pi}{2})$ and the jumps disappear outside this region. To show this more clearly, we plot $\langle n_x(k_y)\rangle$ for $k_z=0$ and $k_z=0.6\pi$ as two examples in Fig. \ref{3DHWfs}(b). The double one-unit-cell jumps driven by $k_y$ indicates that two particles is pumped across the system, as expected for $\mathcal{C}_{k_z}=2$. For comparisons, we also show the results of $\langle n_x(k_y)\rangle$ for $M=4$ without jump for all $k_z$ regime in Figs. \ref{3DHWfs}(c) and \ref{3DHWfs}(d), which is consistent with the expected $\mathcal{C}_{k_z}=0$ when $M>3$ for the band insulators.

\section{Discussion and conclusion}

The properties of the topological Maxwell quasiparticles that are
analogous to the Dirac and Weyl fermions can be further
investigated, for example, the
relativistic wave dynamics with Klein tunneling
\cite{ZhangPRA2012} and Zitterbewegung oscillations \cite{ZLi}, and the unconventional transport properties \cite{Xu2017}.
All of these properties can have unique features. For instance,
there is one oscillation frequency in the Zitterbewegung effects
of the Dirac and Weyl fermions, but there are two different
oscillation frequencies in the Zitterbewegung oscillations of
Maxwell fermions \cite{Shen}. Another idea is to generalize our 2D
model in the presence of the time-reversal symmetry to test if the
quantum spin Hall effect of the 2D Maxwell fermions may occur. In
that case, one can use the cold-atom system to simulate the
quantum spin Hall effect of light \cite{Bliokh}, which has not yet
been realized in experiments. Furthermore, the realization of
Maxwell fermions will open the possibility to many applications,
such as observing the above exotic topological fermions beyond
Dirac and Weyl fermions, simulating the quantum behaviors of
photons in matter, simulating phenomena and solving problems
related to quantum field theory. Therefore, our novel Maxwell
fermions can shed new light on the understanding of Maxwell
equations in quantum field theory and the topological excitations
in condensed matter physics or artificial systems.

In summary, we have systematically explored the topological
Maxwell quasiparticles emerged in Maxwell metals and Maxwell
insulators.  The proposed models can be realized in optical
lattices and the predicted exotic properties of these topological
quasiparticles can be detected in cold-atom experiments.

\acknowledgements{We thank H. J. Zhang and R. B. Liu for useful
discussions. This work was supported by the NKRDP of China (Grant
No. 2016YFA0301803), the NSFC (Grants No. 11474153 and 11604103),
the NSF of Guangdong Province (Grant No. 2016A030313436), and the
Startup Foundation of SCNU.}

\begin{appendix}

\section{Realization of 2D Maxwell points in an optical Lieb lattice}

\begin{figure}[htbp]\centering
\includegraphics[width=7cm]{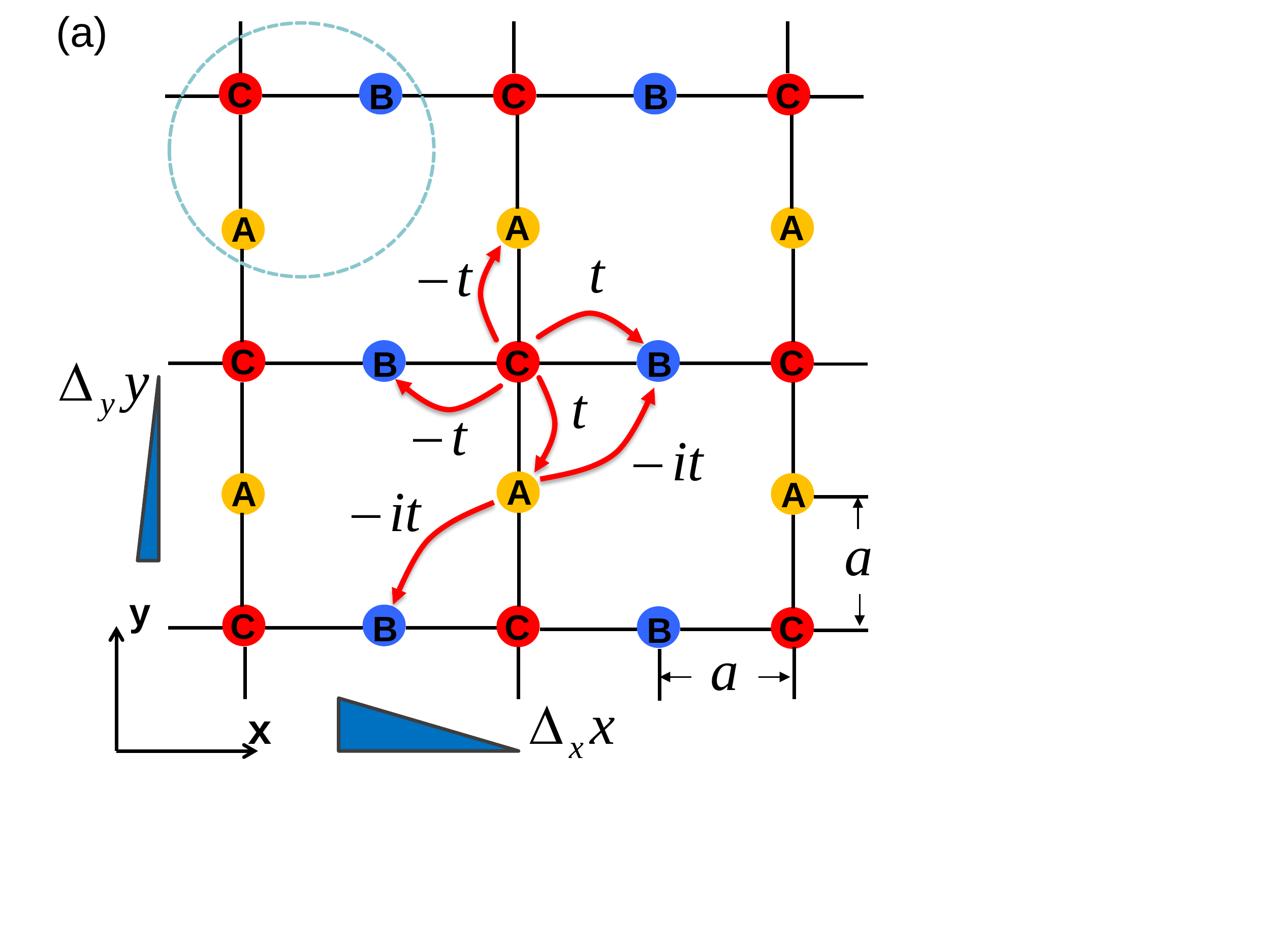}
\includegraphics[width=7cm]{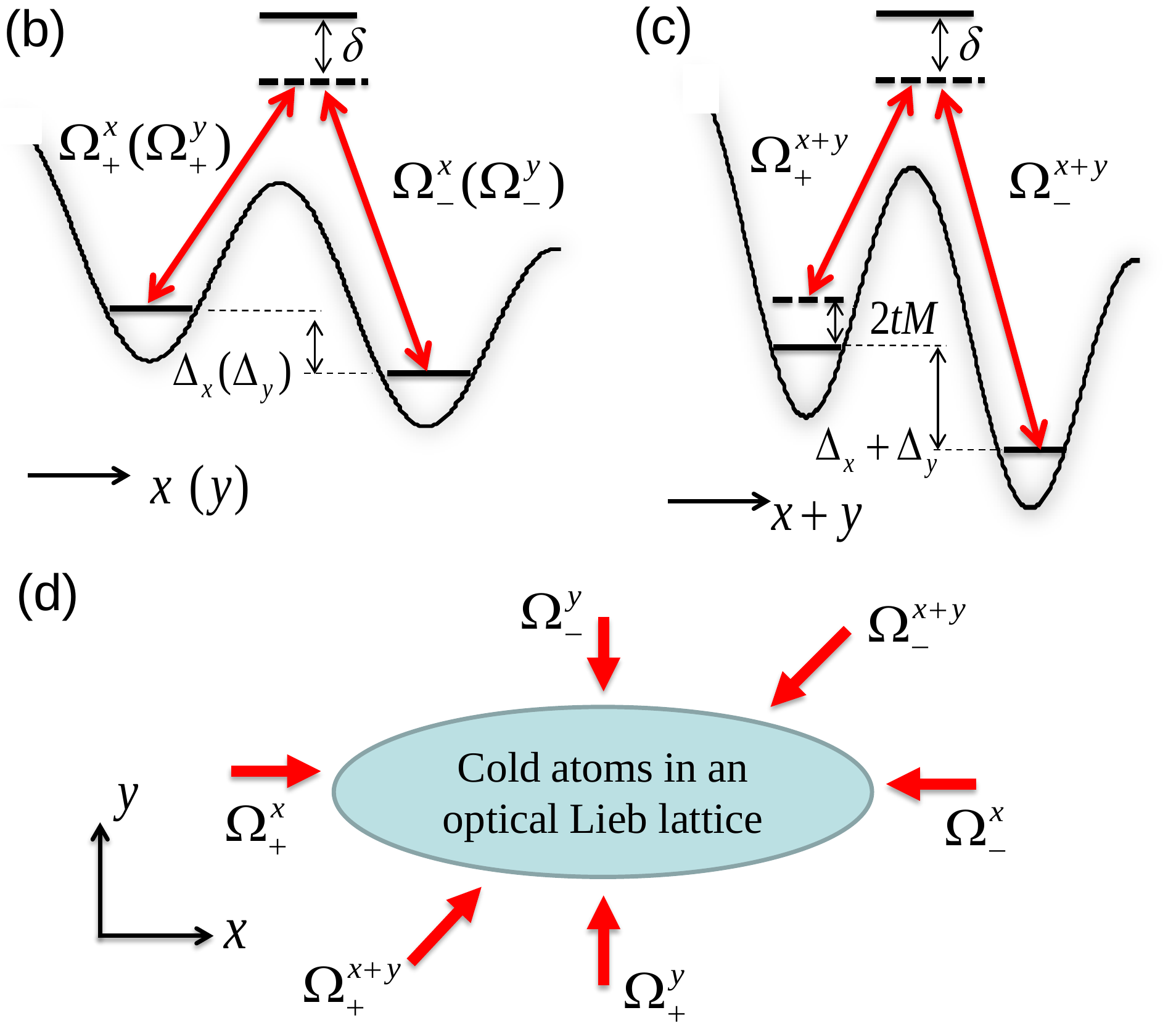}
\caption{(Color online) (a) Schematic diagram of realizing 2D Maxwell fermions in an optical Lieb lattice. A unit cell indicated by dashed line is composed of three sites labeled by $A$, $B$, $C$, with the lattice constant $a$. The three sublattices form the pseudospin-1 basis, and the spin-flip hopping along each direction with the corresponding hopping amplitude is shown. These hopping can be realized by the Raman-assisted hopping method with the help of the linear title potentials $\Delta_{x}x$ and $\Delta_{y}y$ and the application of laser beams, similar as the scheme in the square optical lattice. (b) Two pairs of Raman beams for inducing the desired hopping along $x$ and $y$ directions; (c) A pair of Raman beams for inducing the desired hopping along $x+y$ direction and a two-photon detuning for inducing the constant term $2tM\hat{S}_z$; (d) The total lasers with the corresponding propagation direction. }\label{LiebLattice}
\end{figure}

In this part, we show that the 2D Maxwell points and
the associated Maxwell quasiparticles may be alternatively
realized by using single-component fermionic atoms in optical
lattices with three sublattices, such as an optical Lieb lattice
\cite{Taie}. In experiments, the optical Lieb lattice for cold
atoms has been constructed by superimposing three types of optical
lattices, with the tunable optical potential \cite{Taie}
\begin{equation}
\begin{aligned}
V(x,y)=&-V_{\text{long}}^{x}\cos^2(k_Lx)-V_{\text{long}}^{y}\cos^2(k_Ly)\\
&-V_{\text{short}}^{x}\cos^2(2k_Lx)-V_{\text{short}}^{y}\cos^2(2k_Ly)\\
&-V_{\text{diag}}\cos^2\left[2k_L(x-y)+\frac{\pi}{2}\right].
\end{aligned}
\end{equation}
Here $k_L=2\pi/\lambda$ is a wave number of a long lattice with a depth $V_{\text{long}}$, a short lattice $V_{\text{short}}$ is formed by laser beams at wave length $\lambda/2$, and a diagonal lattice $V_{\text{diag}}$ with the wave number $\sqrt{2}k_L$ is realized by interference of the mutually orthogonal laser beams at $\lambda$ along the $x$ and $y$ directions. The optical Lieb lattice system is shown in Fig. \ref{LiebLattice}(a), with three sublattices $A$, $B$, $C$ forming a unit cell. By tuning the lattice depths $\{V_{\text{long}},V_{\text{short}},V_{\text{diag}}\}$, one can change the energy of the sublattices \cite{Taie}.

In this system, the pseudospin-1 basis are replaced by the three sublattices in a unit cell, and thus the three spin states are given by
$|A\rangle\Longleftrightarrow|\uparrow\rangle, ~|B\rangle\Longleftrightarrow|0\rangle, ~|C\rangle\Longleftrightarrow|\downarrow\rangle$.
In this lattice, the spin-flip hopping $|B\rangle\leftrightarrow|C\rangle$ and $|A\rangle\leftrightarrow|C\rangle$ under the operators $\hat{S}_x$ and $\hat{S}_y$ along the $x$ and $y$ axis become naturally the nearest neighbor hopping in that axis, with the corresponding hopping amplitudes are shown in Fig. \ref{LiebLattice}(a). With the similar Raman-assisted hopping method, the hopping along the $x$ axis and the $y$ axis can be realized by two pairs of laser beams, $\Omega_{\pm}^x=\Omega_0e^{\pm ik_1x}$ and $\Omega_{\pm}^y=\pm\Omega_0e^{\pm ik_1y}$, under the large linear title potentials $\Delta_{x}x$ and $\Delta_{y}y$, respectively, as shown in Fig. \ref{LiebLattice}(b). The detuning in each direction matches the frequency offset of the corresponding Raman beams as we can choose the title energies $\Delta_{x}\approx2.5\Delta_{y}$ with $\Delta_y\gg t_0$ being assumed. In this system, since only one atomic internal state is used in the Raman transitions, then one can address the atoms only through the energy selection without involving the laser polarization \cite{LAT1,LAT2}. Under the two pairs of laser beams, the momenta transferred in the Raman transition along the $x$ and $y$ directions are $\delta \mathbf{k}_1=-2k_1\hat{x}$ and $\delta \mathbf{k}_2=-2k_2\hat{y}$, respectively. Thus the corresponding site-dependent hopping phases along $x$ and $y$ directions are $e^{-2ik_1x}= e^{-2ik_1j_xa}$ and $e^{-2ik_2y}=e^{-2ik_2j_ya}$, with the lattice site index $(j_x,j_y)$. We can choose the parameters $k_1=k_2=\pi/2a$ to induce the hopping phases $e^{-i\pi j_x}=e^{-i\pi j_y}=0,\pi$ staggered along the $x$ and $y$ directions, which lead to the desired hopping $|B\rangle\leftrightarrow|C\rangle$ and $|A\rangle\leftrightarrow|C\rangle$ in the corresponding axis.

The spin-flip hopping $|A\rangle\leftrightarrow|B\rangle$ under
the operator $\hat{S}_z$ in this lattice becomes  next-nearest
neighbor hopping along the $x+y$ or $x-y$ axis, with the
corresponding hopping amplitude along the $x+y$ axis is shown in
Fig. \ref{LiebLattice}(a). This hopping can be achieved by
additional Raman transition by using the third pair lasers
$\Omega_+^{x+y}=\Omega_0e^{ik_3(x+y)}$ and
$\Omega_-^{x+y}=-i\Omega_0e^{-ik_3(x+y)}$ with a different
matching energy $\Delta_x+\Delta_y=3.5\Delta_y$, as shown in Fig.
\ref{LiebLattice}(c). Here a two-photon detuning in the transition
can be used to induce the constant term $2tM\hat{S}_z$, without
adding other coupling beams in this system. We choose the
parameter $k_3=\pi/a$, then the site dependent phase along the
$x+y$ direction can always be reduced to $e^{-2ik_3(j_x+j_y)a}=1$,
such that the hopping constant $-it$ along this direction is
achieved by the two Raman beams. If the hopping
$|A\rangle\leftrightarrow|B\rangle$ along the $x-y$ axis is
wanted, one can also add the Raman transition with the matching
energy $\Delta_x-\Delta_y=1.5\Delta_y$. The laser configuration of
this system is shown in Fig. \ref{LiebLattice}(d). Under these
conditions, the Bloch Hamiltonian of the 2D Maxwell systems  now
becomes
\begin{equation}
\begin{split}
\mathcal{H}(\mathbf{k})=&R_x(\mathbf{k})\hat{S}_x+R_y(\mathbf{k})\hat{S}_y+R_z(\mathbf{k})\hat{S}_z,\\
R_x=&2t\sin{k_x},\\~R_y=&2t\sin{k_y},\\
R_z=&2t[M-\cos(k_x+k_y)].
\end{split}
\end{equation}
Here the spin-1 matrices $\hat{S}_{x,y,z}$ acts on the three
sublattices and the lattice constant $a\equiv1$.  In this case,
one can obtain the Maxwell points and the associated Maxwell
quasiparticles, similar as the case discussed in the main text.
For instance, when the parameter $M=1$, there is a Maxwell point
at $\mathbf{K}=(0,0)$ with the low-energy effective Hamiltonian
$\mathcal{H}_{\text{eff}}(\mathbf{q})\approx
v\emph{q}_x\hat{S}_x+v\emph{q}_y\hat{S}_y$, where $v=2t$ is the
effective speed of light and $\mathbf{q}=\mathbf{k}-\mathbf{K}$.

\section{derivation of the topological invariants}

As we know, the Berry curvature is given by
$\mathbf{F}=\nabla\times\mathbf{A}$, where the Berry connection is
given by
$\mathbf{A}=-i\langle{\psi}|\mathbf{\nabla}{\psi}\rangle$. For the
Bloch Hamiltonian of the 2D model in the main text
$\mathcal{H}=\mathbf{R}(\mathbf{k})\cdot\mathbf{S}$, the Berry
connection $\mathbf{A}=(A_x,A_y,0)$ for the lowest band with the
energy $E=-R$ is given by \cite{He1}
\begin{equation}\label{BC}
A_\mu=-\frac{R_3}{R(R^2-R_3^2)}(R_2\frac{\partial{R_1}}{\partial{k_\mu}}-R_1\frac{\partial{R_2}}{\partial{k_\mu}}).
\end{equation}
The corresponding Berry curvature is $\mathbf{F}=(0,0,F_{xy})$ with $F_{xy}$ being given by
\begin{equation}\label{BF}
\begin{aligned}
F_{xy}
&=\frac{\partial{A_y}}{\partial{k_x}}-\frac{\partial{A_x}}{\partial{k_y}}
\\
&=-\frac{1}{R^3}\varepsilon_{abc}R_{a}\frac{\partial{R_b}}{\partial{k_x}}\frac{\partial{R_c}}{\partial{k_y}}\\
&=-\frac{1}{R^{3}}\mathbf{R}\cdot(\frac{\partial{\mathbf{R}}}{\partial{k_x}}\times\frac{\partial{\mathbf{R}}}{\partial{k_y}}),
\end{aligned}
\end{equation}
where the Bloch vectors are $R_x=2t\sin{k_x}$, $R_y=2t\sin{k_y}$, and $R_z=2t(M-\cos{k_x}-\cos{k_y})$. A straightforward calculation gives the following form
\begin{equation}
F_{xy}=\frac{\cos{k_x}+\cos{k_y}-M\cos k_x\cos k_y}{(\sin^2{k_x}+\sin^2{k_y}+(M-\cos{k_x}-\cos{k_y})^2)^{3/2}}.
\end{equation}
We can thus obtain the Chern number for this band
\begin{equation}
\begin{aligned}
\mathcal{C} &=\frac{1}{2\pi}\oint_Sd\mathbf{k}\cdot\mathbf{F(\mathbf{k})} \\ &=\frac{1}{2\pi}\oint_S{d^2k}F_{xy}\\
&=\left\{\begin{matrix}
2\text{sign}(M), & {(0<|M|<2)}\\
0, &{(|M|>2)h}
\end{matrix}\right.
\end{aligned}
\end{equation}

For $M=\pm2$, we respectively expand the Hamiltonian around $\mathbf{K}_+=(0,0)$ and $\mathbf{K}_-=(\pi,\pi)$, and obtain the low-energy effective Hamiltonian
\begin{equation}
H_{\pm}(\mathbf{q})=\pm(vq_x\hat{S}_x+vq_y\hat{S}_y-2tm\hat{S}_z),
\end{equation}
where $m=2\mp{M}$, $v=2t$, and $\mathbf{q}=\mathbf{k}-\mathbf{K_\pm}$ with $|\mathbf{q}|\ll|\mathbf{k}|$. We can obtain the effective Berry curvature
\begin{equation}
F_{xy}=\pm\frac{m}{(q^2+m^2)^{3/2}},
\end{equation}
where $q=\sqrt{q_x^2+q_y^2}$. Thus the Berry phase $\gamma$ integrated around the Maxwell point $\mathbf{K}_\pm$ for the Fermi surface  can be derived by
\begin{equation}
\begin{aligned}
\gamma=\oint_{FS}d\mathbf{k}\cdot\mathbf{A(\mathbf{k})}&=\pm\oint_{FS}{d^2q\frac{m}{(q^2+m^2)^{3/2}}}\\
&=\pm\int_0^{2\pi}d\theta\int_{0}^{k_F}\frac{m}{(q^2+m^2)^{3/2}}qdq\\
&=\pm{2\pi\int_{0}^{k_F}\frac{m}{(q^2+m^2)^{3/2}}qdq},
\end{aligned}
\end{equation}
where $k_F$ is the Fermi momentum and the parameter $m\rightarrow0$. Let $q=m\tan{\varphi}$, then we have $1+\tan^2\varphi=\sec^2\varphi$ and $dq=m\sec^2\varphi{d\varphi}$. Substituting these relationships into the above equation, we obtain $\gamma$ as a function of $m$:
\begin{equation}
\begin{split}
\gamma&=\pm2\pi\int_{0}^{k_F}\frac{m}{(q^2+m^2)^{3/2}}qdq\\
&=\pm2\pi\int_{0}^{\varphi_F}\frac{m^2\tan{\varphi}}{m^3\sec^3{\varphi}}m\sec^2{\varphi}d\varphi\\
&=\pm2\pi\int_0^{\varphi_F}\sin{\varphi}d\varphi \\
&=\pm2\pi(1-\frac{m}{\sqrt{k^2_F+m^2}}).
\end{split}
\end{equation}
Thus for $m=0$, we obtain $\gamma=\pm2\pi$ for $M=\pm2$.

For $M=0$, we respectively expand the Hamiltonian around
$\mathbf{K}_{(0,\pi)}=(0,\pi)$ and $\mathbf{K}_{(\pi,0)}=(\pi,0)$,
and obtain the low-energy effective Hamiltonian
\begin{equation}
\mathcal{H}_0(\mathbf{q})=\pm({vq_x\hat{S}_x-vq_y\hat{S}_y+2tm_0\hat{S}_z})
\end{equation}
where $m_0=\pm{M}$ in this case, and $\mathbf{q}=\mathbf{k}-\mathbf{K}_{(0,\pi)/(\pi,0)}$, $|\mathbf{q}|\ll|\mathbf{k}|$. We can obtain the Berry connection
\begin{equation}
\begin{split}
A_x &=\pm\frac{m_0q_y}{q^2\sqrt{q^2+m_0^2}}, \\ ~A_y &
=\pm\frac{m_0q_x}{q^2\sqrt{q^2+m_0^2}}. \end{split}
\end{equation}
Thus the Berry phase $\gamma$ integrated around the Maxwell point $\mathbf{K}_{(0,\pi)/(\pi,0)}$ for Fermi surface can be derived as
\begin{equation}
\begin{split}
\gamma&=\oint_{FS}d\mathbf{k}\cdot\mathbf{A(\mathbf{k})}=\int_0^{2\pi}k_Fd\theta{A_\theta}\\
&=\int_0^{2\pi}k_Fd\theta(A_y\frac{q_x}{k_F}-A_x\frac{q_y}{k_F})\\
&=\pm\frac{m_0}{\sqrt{k_F^2+m_0^2}}\int_0^{2\pi}(\cos^2\theta-\sin^2\theta)d\theta\\
&=\pm\frac{m_0}{\sqrt{k_F^2+m_0^2}}\int_0^{2\pi}d\theta{\cos(2\theta)}
\\
& =0,
\end{split}
\end{equation}
where we have used the relationships $q_x=k_F\cos\theta$ and $q_y=k_F\sin\theta$.

Below we calculate the monopole charge of the 3D Maxwell points.
For the effective Hamiltonian around the Maxwell point $\mathbf{M}_+=(0,0,\frac{\pi}{2})$ in the main text, we use a more simplify form
\begin{equation}
H(\mathbf{q})=q_x\hat{S}_x+q_y\hat{S}_y+q_z\hat{S}_z
\end{equation}
and then we obtain the three energies are given by $\epsilon_\pm=\pm|\mathbf{q}|$, $\epsilon_0=0$. Going to polar coordinates $(q_x,q_y,q_z)=|\mathbf{q}|(\sin{\theta}\cos\varphi,\sin{\theta}\sin{\varphi},\cos{\theta})$, the corresponding eigenfunctions are
\begin{equation}
\begin{split}
\psi_\pm &=\frac{1}{\sqrt{2}}\begin{pmatrix}\pm{i\sin{\varphi}}-\cos{\theta}\cos{\varphi}\\
\mp{i\cos{\varphi}-\cos{\theta}\sin{\varphi}}\\ \sin{\theta}
\end{pmatrix}, \\
~ \psi_0&=\begin{pmatrix} \sin{\theta}\cos{\varphi}\\
\sin{\theta}\sin{\varphi}\\ \cos{\theta}\end{pmatrix}.
\end{split}
\end{equation}
The relationship for Berry curvature in different coordinates
\begin{equation}
\begin{aligned}
F^{i}&=\epsilon_{ijk}F_{jk}=\epsilon_{ijk}F_{\theta\varphi}\frac{\partial(\theta,\varphi)}{\partial(R_j,R_k)},\\
F_{\theta\varphi} &=\partial_\theta{A_\varphi}-\partial_\varphi{A_\theta},\\\
\end{aligned}
\end{equation}
where $A_\theta=i\langle{\psi}|\nabla_\theta|{\psi}\rangle$, and
$ A_\varphi =i\langle{\psi}|\nabla_\varphi|{\psi}\rangle$. To
each of these eigenfunctions, the associated U(1) Berry curvature
is given by
\begin{equation}
\begin{aligned}
\mathbf{F(k)}_\pm&=\nabla\times i\langle{\psi_\pm}|\nabla|{\psi_\pm}\rangle=\mp\frac{\mathbf{k}}{|\mathbf{k}|^3},\\
\mathbf{F(k)}_0&=\nabla\times i\langle{\psi_0}|\nabla|{\psi_0}\rangle=0
\end{aligned}
\end{equation}
The integral of $\mathbf{F(k)}_\pm$ over any surface enclosing the Maxwell point $\mathbf{M_+}$ is
\begin{equation}
\mathcal{C}_\pm=\frac{1}{2\pi}\oint_Sd\mathbf{k}\cdot\mathbf{F(k)}_\pm=\mp2
\end{equation}
Therefore, the monopole charge of the Maxwell point $\mathbf{M}_+$ for the lowest band is $\mathcal{C}=+2$. The monopole charge of the other Maxwell point $\mathbf{M}_-$ for the lowest band is obtained as $\mathcal{C}=-2$.
Fortunately, we can obtain the formula of Berry curvature $\mathbf{F(k)}$ for the lowest band by using (\ref{BC}) and (\ref{BF}), which are given by
\begin{eqnarray} \nonumber
F^x&=&{\sin{k_x}\cos{k_y}\sin{k_z}}/{G(\mathbf{k})},\\
F^y&=&{\cos{k_x}\sin{k_y}\sin{k_z}}/{G(\mathbf{k})},\\ \nonumber
F^z&=&({M\eta-\cos{k_x}-\cos{k_y}-\eta\cos{k_z}})/{G(\mathbf{k})},
\end{eqnarray}
where $\eta=\cos{k_x}\cos{k_y}$ and
$G(\mathbf{k})=[\sin^2{k_x}+\sin^2{k_y}+(M-\cos{k_x}-\cos{k_y}-\cos{k_z})^2]^\frac{3}{2}$.
Thus one can easily get the Berry curvature $\mathbf{F(q)}$ around
the Maxwell points.

\end{appendix}





\end{document}